\documentclass[fleqn,usenatbib]{mnras}

\usepackage{newtxtext,newtxmath}

\usepackage[T1]{fontenc}

\DeclareRobustCommand{\VAN}[3]{#2}
\let\VANthebibliography\thebibliography
\def\thebibliography{\DeclareRobustCommand{\VAN}[3]{##3}\VANthebibliography}

\usepackage{graphicx}	
\usepackage{amsmath}	

\defcitealias{Marcus2010}{M10}
\defcitealias{Reinhardt2022}{R22}
\defcitealias{Dou2024}{D24}

\title[Maximum Density]{The Maximum Density of a Collisionally-Produced Planet is A Function of its Mass and Orbital Period}

\author[M. Brady et al.]{
Madison Brady,$^{1}$\thanks{E-mail: bradym27@msu.edu}
Darryl Z. Seligman,$^{1}$
\\
$^{1}$ Dept. of Physics and Astronomy, Michigan State University, East Lansing, MI 48824, USA \\
}

\date{Accepted XXX. Received YYY; in original form ZZZ}

\pubyear{2026}

\begin{document}
\label{firstpage}
\pagerange{\pageref{firstpage}--\pageref{lastpage}}
\maketitle

\begin{abstract}

There are many different theoretical explanations for the formation of high-density Mercury-like planets, but concrete evidence for any of these formation mechanisms remains elusive. A popular explanation for dense planets is the collisional hypothesis, which states that iron-rich planets can be formed as the products of high-energy, mantle-stripping impacts.  Planetesimal collision simulations predict that higher-velocity collisions can form higher-density planets.  Motivated by the characteristics of the high-density, short-period ($P\,\approx\,0.3$d) GJ\,367\,b, we study the results of previously-published smoothed-particle hydrodynamics (SPH) simulations on exoplanet collisions, combining these with models describing the likely collision velocities of these objects, to investigate the relationship between the core mass fractions (CMFs) of exoplanets, their masses, and their orbital periods.   We predict that collisionally-produced super-Mercuries should be more common (and more dense) at low masses and short orbital periods.  This correlation may enable us to pinpoint the formation mechanism of super-Mercuries as the population of observed targets grows.  Afterwards, we connect our hypothesis to the observed Mercury-like population of high-density exoplanets, and find that GJ\,367\,b is the best exoplanetary candidate for collisional formation.

\end{abstract}




\section{Introduction}

The terrestrial objects in our solar system have a broad range of densities, from the iron-rich Mercury to the light and rocky moon.  However, the two largest terrestrial planets (Venus and Earth) have bulk densities that imply similar internal compositions.  Each of these planets is consistent with roughly one-third of their mass being stored in an iron-rich core while the remaining two-thirds of their mass are composed of a rocky mantle.  The similarity between the inferred core mass fractions (CMFs) of these larger planets may indicate that some additional mechanism must be called upon to explain planets with more extreme CMFs.

There are several potential explanations for the high \citep[$\approx$0.7;][]{Hauck2013} CMF of Mercury.  One is that it formed collisionally, as the result of an impact on a mostly-differentiated planet that stripped away its mantle \citep{Benz1988}.  However, this explanation has been disfavored more recently, as N-body simulations have shown that collisions energetic enough to strip its mantle from its core would be rare \citep{Franco2022}, though it may be possible in some specific scenarios \citep{Franco2025}.  

In addition, MESSENGER probe observations of Mercury have shown the presence of various moderately volatile elements on its surface \citep{Peplowski2011}, which would have been unlikely to survive a high-energy impact.  Therefore, several papers have theorized that Mercury has a primordially high CMF, which could be caused by some mechanism which causes iron to become concentrated in the inner disk.  There are many different proposed processes that may cause this concentration.  \cite{Kruss2018} and \cite{Kruss2020} theorized that aggregates of iron particles could be caused by host-induced magnetic fields in the inner disk.  \cite{Ebel2011} found that the conditions in the inner disk resulted in more volatile silicon and less volatile iron, resulting in the formation of planetesimals with a higher Fe/Si ratio. \cite{Johansen2022} stated that the high surface tension of iron may lead it to form homogeneous pebbles in the presence of disk heating events.  Furthermore, works such as \cite{Ebel2018} have described mechanisms that enable Mercury to retain the high volatile abundances observed by MESSENGER even after massive impacts.  However, there are currently few observational clues as to which process correctly describes Mercury's formation.

We may be able to learn more about how planets form if we expand our sample to include exoplanets.  The majority of terrestrial exoplanets appear to have CMFs roughly consistent with that of the Earth \citep[see, e.g.,][]{Brinkman2025}, indicating that planets with a CMF$\approx$0.3 are a common outcome of planet formation.  However, the discovery of high-density exoplanets outside of the solar system, such as GJ\,367\,b \citep{Goffo2023} and Kepler-107\,c \citep{Bonomo2019}, give us another opportunity to study the formation mechanism of super-dense planets.  Beyond the impact and primordial concentration hypotheses, there have been several other explanations developed to explain the formation of high-density planets.  Works such as \cite{Adibekyan2021} and \cite{Plotnykov2025} theorized that high-density planets are the result of formation around metal-rich stars, while \cite{Mocquet2014} presented the hypothesis that high-density planets may be the compressed cores of giant planets that recently lost their envelopes.  It is necessary to quantify the observational implications of these different formation mechanisms if we want to understand which is responsible for super-Mercury formation.

The formation of high-density exoplanets via collisions have been studied extensively using smoothed-particle hydrodynamics (SPH) simulations.  \cite{Marcus2010} attempted to quantify the upper limit to the CMFs of planets utilizing the results of the impact simulations in \cite{Marcus2009}. They fit a formula that related the mass and iron fraction of the largest impact-created fragment to the impact velocity, which enabled them to create a mass-radius curve as a function of velocity.  In general, they found that lower-velocity collisions (which are less energetic) result in less mantle stripping and therefore lower CMF remnants.  They then produced a ``maximum collision-stripping'' upper limit to the mass of rocky planets as a function of radius, assuming a high impact velocity of $v_\mathrm{imp}\,=\,80$\,km\,s$^{-1}$, which is much higher than the anticipated impact velocity at 1 AU around a solar-type star.  In general, their results predicted that it is more difficult to strip the mantles off of higher-mass planets, resulting in high-mass planets having a lower upper limit to their CMFs than low-mass planets.  Therefore, if super-Mercuries are formed collisionally, the densest objects will be found at the lowest masses.

Many later studies have iterated upon the work from \cite{Marcus2010}.  For example, \cite{Reinhardt2022}, hereafter referred to as \citetalias{Reinhardt2022}, expanded the range of planet masses and impact velocities explored, as well as corrected some calculation errors.  In addition, \cite{Dou2024}, hereafter referred to as \citetalias{Dou2024}, performed another suite of simulations, including grazing hit-and-run impacts with nonzero impact parameters.  Such impacts are likely far more common than the head-on impacts described in other works.  Both of these studies recovered similar qualitative results as \cite{Marcus2010}, finding that higher-velocity impacts produce higher-CMF planets and that the highest-CMF planets have the lowest masses. 

While all of these studies have recovered similar relationships between impact velocity and remnant CMF, it is necessary to consider which impact velocities are realistic for a given planet.  Many studies \citep[such as][]{Marcus2010, Franco2025} follow the conclusions of \cite{Agnor1999} and assume that most collisions occur at some multiple of the mutual escape velocity of the two impacting planetesimals.  While the mutual escape velocity provides a floor on the collision velocity \citep[see, e.g.,][]{Moorhead2020}, impact velocities can be much higher in regions in the planetary disk where the planets are orbiting more rapidly.  This can result in extremely dramatic impacts at short orbital periods around high-mass stars.  Describing the impact velocity in terms of the mutual escape velocity does not capture this period dependence.  

Recently, \cite{Cambioni2025} performed a series of simulations of compact super-Earths with different sizes and degrees of excitation, using machine learning on the SPH impact database by \cite{Emsnehuber2024} (which only considered head-on collisions) to model the impact results.  They found that erosive hit-and-run collisions can produce super-Earths similar to the observed high-density exoplanet population, but concluded that such bodies have a $\approx 1\%$ occurrence rate, which may be too low to explain the observed densities of super-Earths, concluding that either our sample of observed exoplanets is biased or that some other mechanism produces high-density super-Earths.  However, they did find that smaller planets (such as sub-Earths) are more likely to form metal-rich. In this work, we performed an independent analysis to study how high-density planets form collisionally.

In this paper, we will estimate the anticipated impact velocity of planetesimals in the protoplanetary disk as a function of orbital period and host star mass using both simulations and analytical techniques (Section~\ref{sec:methods}).  We will then combine these impact velocities with the fit equations from \citetalias{Reinhardt2022} and \citetalias{Dou2024} to estimate a more realistic range of planet CMFs as a function of their orbital periods.  We will then compare these results to the observed CMFs of super-dense planets and determine whether or not they are plausible given our results (Section~\ref{sec:discussion}).  We conclude in Section~\ref{sec:summary_conclusions}.

\section{Methods}
\label{sec:methods}

There are many studies that have reported the results of SPH simulations of planetary collisions for the purposes of creating straightforward fitting functions that can be used to predict the outcome of planetesimal collisions.  We will elaborate upon these models by calculating the likely velocities and impact parameters of these collisions, and use those to estimate more realistic limits on exoplanet densities as a function of orbital period, host star mass, and planetary mass.

We focus specifically on the SPH results from the simulations of \citetalias{Reinhardt2022} and \citetalias{Dou2024}, as they are both recent studies that fit equations to the CMFs of their post-impact planets.  These studies are relatively similar in terms of the range of planetary masses they examined (\citetalias{Reinhardt2022} examines impactor masses between 1 --- 20 M$_\oplus$, while \citetalias{Dou2024} goes from 0.06 --- 20 M$_\oplus$).  

Neither simulation allows for re-accretion of impact debris onto the surface of the post-impact planetesimal.  \citetalias{Reinhardt2022} chose to not model this process because there are several mechanisms that may prevent ejecta re-accretion.  As an example, \cite{Spalding2020} showed that stellar winds might cause the rapid semi-major axis evolution of the collisional debris, removing it from the orbit of the newly-changed planet and preventing it from re-accreting.  As we anticipate that stellar winds are more dramatic early on in the system's lifetime, when planets are likely forming, this mechanism may be relevant to collisionally-produced planets.  In addition, \cite{Benz2008} showed that Poynting-Robertson drag may also result in the removal of impact debris from the system.  \citetalias{Reinhardt2022} also theorized that the small Hill radii (and enhanced tidal forces) of close-in planets may also play a role in limiting re-accretion.  Meanwhile, \citetalias{Dou2024} did not account for re-accretion, as they intended on finding a lower limit on planetary radii, though they also noted the possibility of ejecta loss via solar winds.  Therefore, we do not consider the lack of re-accretion in \citetalias{Dou2024} or \citetalias{Reinhardt2022} to be an obvious cause for concern, though we note that the potential re-accretion of debris could lower the maximum collisionally-induced CMFs for these models.

With regards to the differences between these two studies, \citetalias{Reinhardt2022} performed their fits with a broad range of different impactor-to-target mass ratios, while \citetalias{Dou2024} assumed that all collisions were equal-mass.  Meanwhile, \citetalias{Dou2024} calculated the results of impacts with nonzero impact parameters, which is a more realistic prescription than \citetalias{Reinhardt2022}, which assumed all collisions were head-on.  Given these differences, we will examine both models and compare their results.

\subsection{Results From \citetalias{Reinhardt2022}}
\label{ssec:r22}

Using the results from SPH simulations, \citetalias{Reinhardt2022} calculated the mass and iron fraction of the largest fragment mass as a function of $Q_R$/$Q^*_{RD}$, where $Q_R$ is the specific impact energy of the collision and $Q^*_{RD}$ is the catastrophic destruction threshold, which is defined as the collision energy necessary to strip half of the colliding mass from the system.

$Q_R$ is defined as:

\begin{equation}
    Q_R =\frac{1}{2} \frac{m_t m_i }{(m_t+m_i)^2}v_\mathrm{imp}^2,
\end{equation}

\noindent where $m_t$ is the target mass, $m_i$ is the impactor mass, and $v_\mathrm{imp}$ is the impact velocity.  The critical specific energy $Q^*_{RD}$ was fit in \citetalias{Reinhardt2022} and has the following form:

\begin{equation}
\label{eqn:qrd}
    Q^*_{RD} = q\times   R_{C1}^{3 \mu } v_\mathrm{imp}^{2-3 \mu},
\end{equation}

\noindent where $q$ and $\mu$ are fit constants and $R_{C1}$ is equal to:

\begin{equation}
    R_{C1} = \bigg(\frac{3(m_t+m_i)}{4 \pi \rho_w} \bigg)^{1/3},
\end{equation}

\noindent where $\rho_w$ is the density of water. 

\citetalias{Reinhardt2022} then fit a series of equations to calculate the mass of the largest fragment $m_{lr}$ and remnant core mass fraction $CMF_{lr}$ as a function of $Q_R/Q_{RD}^*$.  We will not reproduce them to save space, but refer the readers to Equations 4-7 in \citetalias{Reinhardt2022} for reference.

If we define the mass ratio $\gamma$ such that $m_i = \gamma m_t$, we can differentiate the $CMF_{lr}$ function with respect to $\gamma$ to find the mass ratio that produces the planets with the highest core-mass fraction values. \citetalias{Reinhardt2022} found that $CMF_{lr}$ was a piecewise function that monotonically increased with $Q_R/Q^*_{RD}$, meaning that we can find the maximum value of $CMF_{lr}$ at a fixed velocity by finding the maximum value of $Q_R/Q^*_{RD}$.  This ratio can be written as:

\begin{equation}
 \frac{Q_R}{Q^*_{RD}} =  \frac{1}{2}\frac{m_t m_i }{ (m_t+m_i)^{2+\mu}  }\frac{v_\mathrm{imp}^{3\mu} (4 \pi \rho_w)^\mu}{3^{\mu}q},
\end{equation}

\noindent which can be written in terms of $m_t$ and $\gamma$:

\begin{equation}
 \frac{Q_R}{Q^*_{RD}} =  \frac{1}{2}\frac{v_\mathrm{imp}^{3\mu} (4 \pi \rho_w)^\mu }{ 3^\mu q m_t^{\mu} } \gamma (\gamma+1)^{-2-\mu}.
\end{equation}

We can then differentiate with respect to $\gamma$ and set the equation equal to zero to find the maximum value:

\begin{equation}
 0 =  (\gamma+1)^{-2-\mu} + \gamma(-2-\mu)(\gamma+1)^{-3-\mu}.
\end{equation}

We can eliminate the $\gamma=-1$ solution, as it is unphysical (referring to a planet with a negative mass), and find:

\begin{equation}
\gamma = \frac{1}{1+\mu}.
\end{equation}

\begin{figure}
    \centering
    \includegraphics[width=1\linewidth]{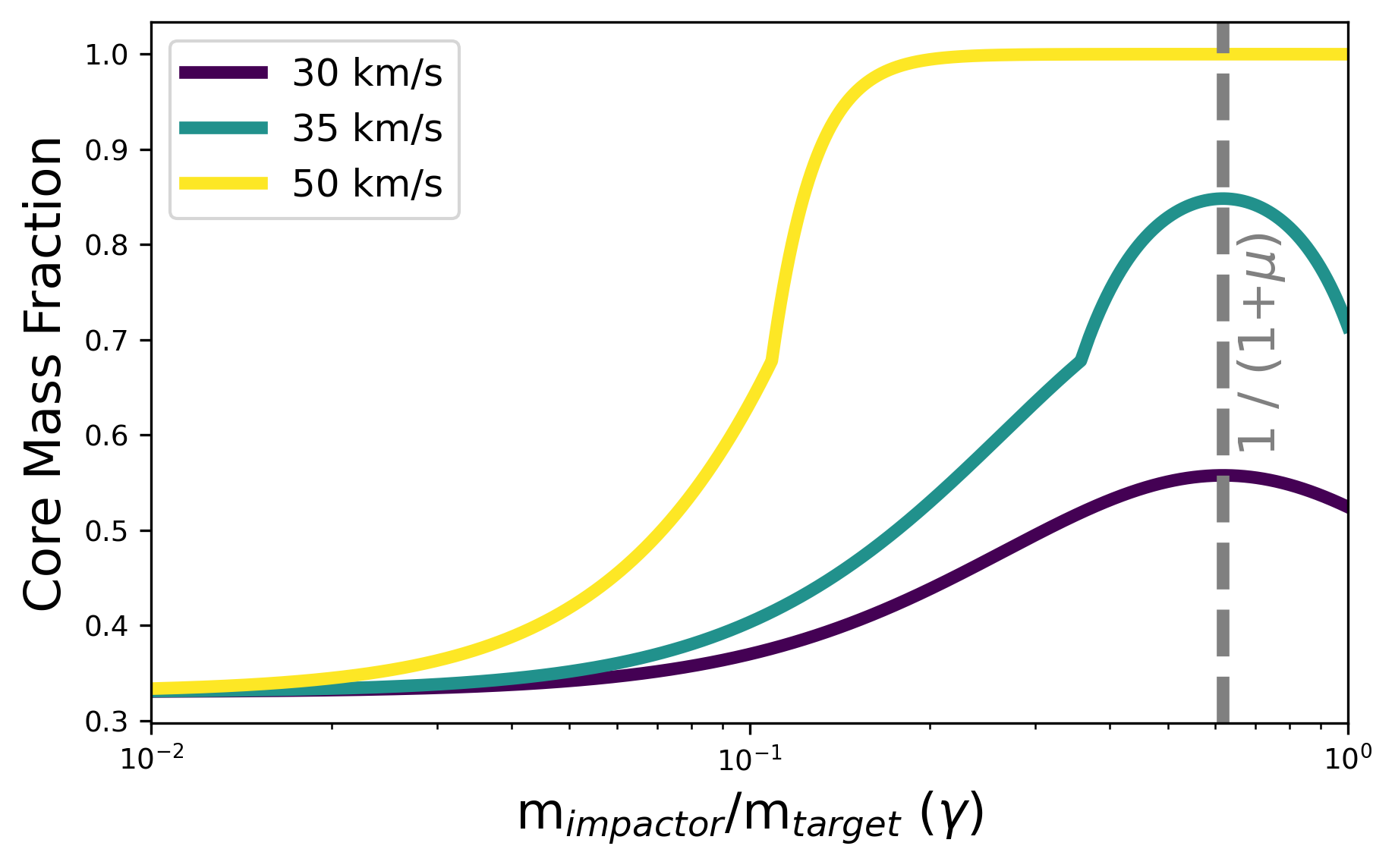}
    \caption{The calculated core mass fraction of a 1$M_\oplus$ planet as a function of $\gamma$ for a variety of different impact velocities, using the equations from \citetalias{Reinhardt2022}.  We highlight the value of $\gamma$ that maximizes the core mass fraction with a dashed gray line.  The core mass fraction is maximized when $\gamma\,\approx\,0.6$, but $\gamma\,=\,1$ appears to provide a reasonable substitute in most cases.}
    \label{fig:gamma_z}
\end{figure}

Therefore, $CMF_{lr}$ is maximized when $\gamma = \frac{1}{1+\mu}$, where $\mu$ is the exponent in Equation~\ref{eqn:qrd}, which \citetalias{Reinhardt2022} fit to be equal to $\mu\,=\,0.6164$.  With this equation, we find that $CMF_{lr}$ is maximized when $\gamma\,=\,0.6187$ (see Figure~\ref{fig:gamma_z}).

Interestingly enough, this means that $\gamma$ is not maximized in these fits with an equal-mass impactor and target, but is instead maximized when the impactor is slightly smaller than the target.  However, we note that, in most cases, the difference in $CMF_{lr}$ when assuming $\gamma\,=\,1$ vs $\gamma\,=\,0.6$ collisions is usually less than 10\% (reaching a maximum value around 20\%), so this change is relatively minor.

\begin{figure}
    \centering
    \includegraphics[width=1\linewidth]{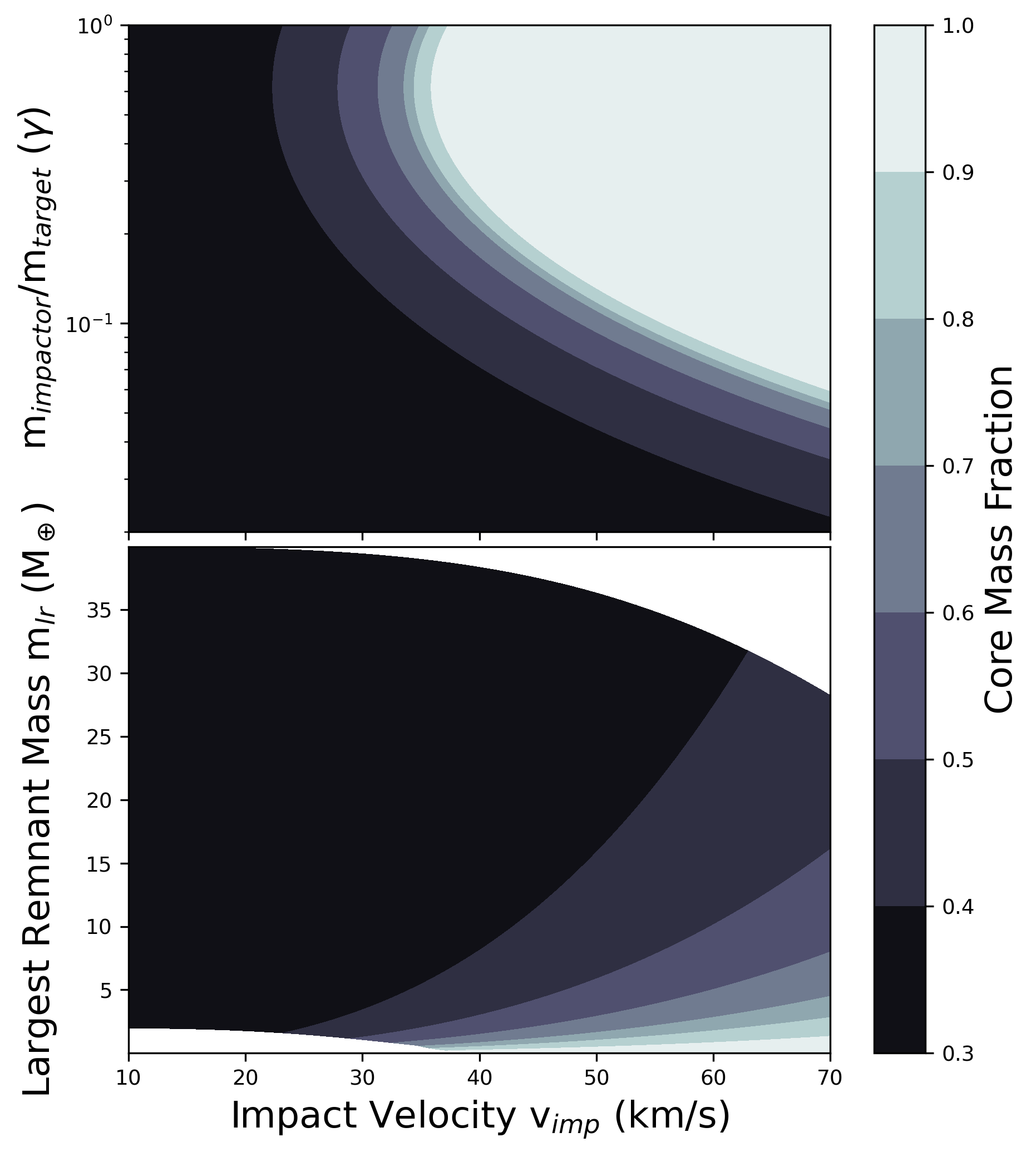}
    \caption{\textit{Top:} The core mass fraction $CMF_{lr}$ (calculated using equations from \citetalias{Reinhardt2022}) as a function of the impact velocity $v_\mathrm{imp}$ and mass ratio $\gamma$ for a target mass of $m_t\,=\,$1\,M$_\oplus$.  \textit{Bottom:} $CMF_{lr}$ as a function of $v_\mathrm{imp}$ and the mass of the largest remnant $m_{lr}$ for target masses between 1--20\,$M_\oplus$.  We assumed $\gamma\,=\,1$ for these calculations. In general, it is clear that high-velocity impacts where $\gamma\,\approx\,1$ produce planets with the highest $CMF_{lr}$ values.  In addition, it is easier (as in, it requires less energetic impacts) to form low-mass planets with high $CMF_{lr}$ values than it is to form super-dense planets with larger masses. }
    \label{fig:vimp_gamma_cmf}
\end{figure}

Figure~\ref{fig:vimp_gamma_cmf} shows how the core mass fraction of the largest collisional remnant is a function of $\gamma$, the collision velocity, and the actual mass of the remnant. In general, we note that higher-velocity collisions tend to result in fragments with larger core-mass fractions.  This makes sense, as high-velocity collisions between similarly-sized bodies are more energetic and therefore more capable of stripping a planetesimal's mantle.  In addition, it is easier (as in, it requires lower velocities) to create high-CMF planets at low masses than it is at higher masses, likely because the mantles of these planets are easier to strip.  Finally, as we calculated earlier, the highest $CMF_{lr}$ values are achieved when $\gamma\,\approx\,0.6$.  It also appears to be extremely difficult to form high-CMF planets when $\gamma$ is less than about 0.1, indicating that high-density planets are most likely to be formed as a result of collisions of bodies with similar masses (at least according to the \citetalias{Reinhardt2022} simulations).

\subsection{Results from \citetalias{Dou2024}}

\citetalias{Dou2024} and \cite{Dou2024b} also provided a prescription for modeling the results of impacts between planetesimals. However, their prescription is different from that of \citetalias{Reinhardt2022} in several key ways: 1) they have a different equation for the $Q_{RD}^*$ disruption criterion, 2) they fit their results with formulae with different functional forms, and 3) they consider oblique (in addition to head-on) collisions.

One of the first differences is the form they use for $Q^*_{RD}$.  Following \cite{Leinhardt2012}, they fit the function:

\begin{equation}
    Q^*_{RD,\,\gamma=1} = c^* \pi \rho_w G R_{C1}^2
\end{equation}

\noindent where $c^*$ is a constant.  While \citetalias{Dou2024} does not explicitly describe how this changes with $\gamma$, \cite{Leinhardt2012} states:

\begin{equation}
    Q^*_{RD} = Q^*_{RD,\,\gamma=1} \bigg(\frac{(\gamma-1)^2}{4\gamma} \bigg)^{2/(3\mu) - 1}.
\end{equation}

By differentiating this function, we find that the ratio $Q_R/Q^*_{RD}$ is maximized when $\gamma\,=\,1$, unlike the observations from \citetalias{Reinhardt2022}.  We also note that $Q^*_{RD}$ in \citetalias{Dou2024} (and its related references) has a different radius and velocity dependence than that from \citetalias{Reinhardt2022}:

\begin{equation}
\begin{aligned}
    Q^*_{RD,\,\mathrm{R}} \propto R_{C1}^{3\mu} v_\mathrm{imp}^{2-3\mu} \\
    Q^*_{RD,\,\mathrm{D}} \propto R_{C1}^{2}
\end{aligned}
\end{equation}

However, if we substitute $\mu\,=\,0.6164$ from \citetalias{Reinhardt2022}:

\begin{equation}
    Q^*_{RD,\,\mathrm{R}} \propto R_{C1}^{1.85} v_\mathrm{imp}^{0.15} .
\end{equation}

Therefore, the velocity dependence on $Q^*_{RD}$ appears to be weak in \citetalias{Reinhardt2022}.  These two functional forms further agree when considering the results from \cite{Stewart2009}, which showed a linear relationship between the velocity necessary to cause catastrophic disruption and $R_{C1}$, which is foundational to the assumptions that constructed the $Q_{RD}^*$ formulations in \cite{Leinhardt2012} and \citetalias{Dou2024}.  However, while this difference is subtle, it does result in \citetalias{Dou2024} predicting a lower catastrophic disruption criterion at lower velocities and a higher criterion at high velocities, which may result in a higher inferred core mass fraction limit at low velocities (and vice versa). 

\begin{figure}
    \centering
    \includegraphics[width=1\linewidth]{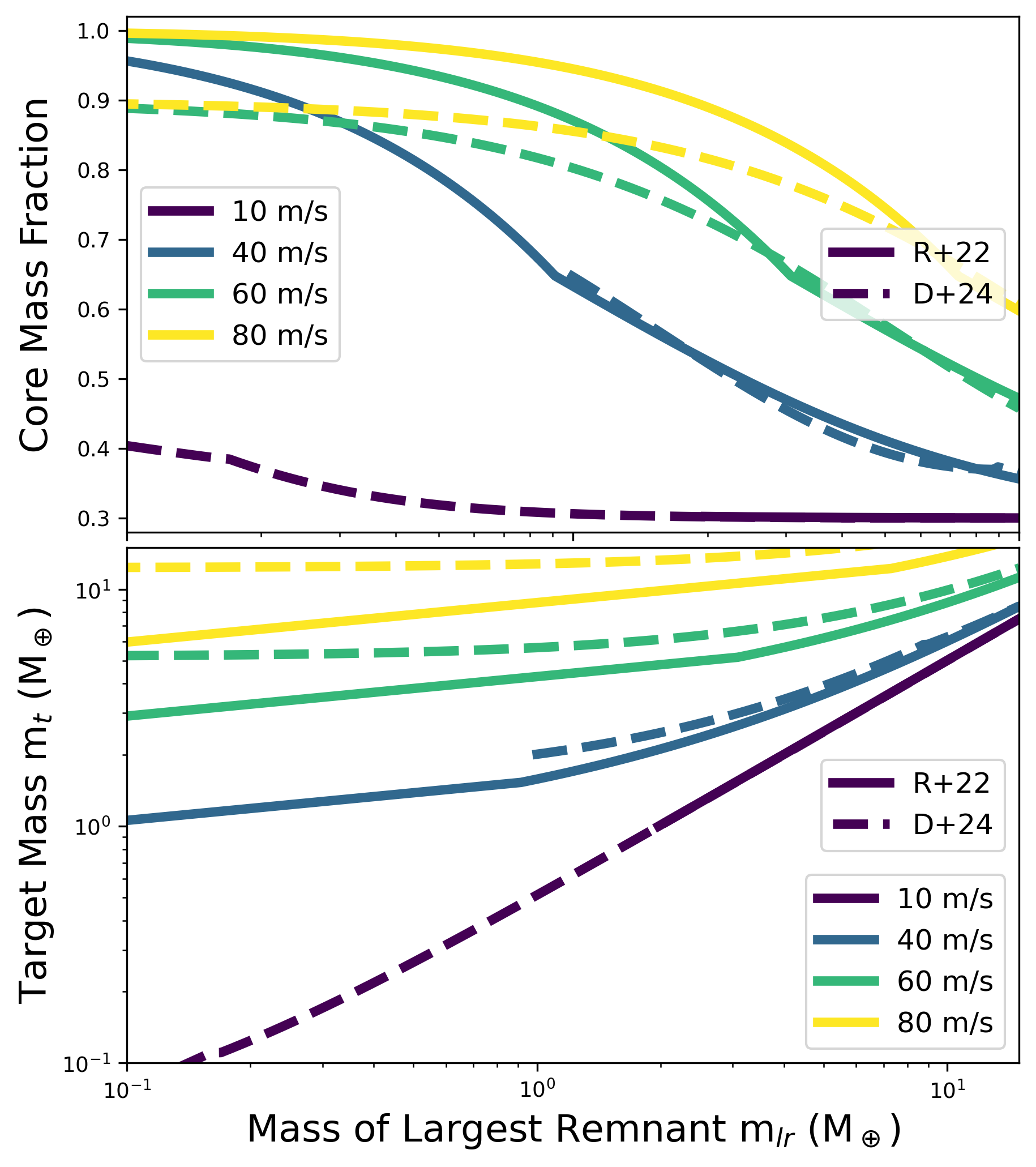}
    \caption{A comparison between the \citetalias{Reinhardt2022} (solid lines) and \citetalias{Dou2024} (dashed lines) results at various impact velocities assuming head-on impacts and  $\gamma\,=\,1$.   \textit{Top:} The mass of the largest remnant $m_{lr}$ vs. the CMF.  \textit{Bottom:} The target mass $m_t$ vs. the mass of the largest remnant $m_{lr}$.}
    \label{fig:rd_comp}
\end{figure}

\citetalias{Dou2024} also calculated the mass and CMF of the largest fragment as a function of the $Q_R/Q_{RD}^*$ ratio. To save space, we refer the reader to equations 4-15 in \citetalias{Dou2024} and not re-iterate them here.  We show the results of their fits for $\gamma\,=\,1$ head-on collisions in Figure~\ref{fig:rd_comp}, compared to those from \citetalias{Reinhardt2022}. 

One key difference in \citetalias{Dou2024} is that they fit separate relations for $CMF_{lr}$ and $m_{lr}$ based on the target mass in their head-on simulations, while \citetalias{Reinhardt2022} did not.  In addition, the authors did not track simulations where $m_{lr}/(m_t+m_i)\,<0.1$ given the resolutions of their simulations, so their results are incomplete in this regime.  They found that collisions for low target masses when $Q_R/Q_{RD}^*\,>\,1.2$ tended to cause dramatic fragmentation that resulted in $m_{lr}/(m_t+m_i)\,<0.1$, so they recommended against using their prescriptions for low-mass (defined in their case as $m_t\,<\,2$M$_\oplus$) impacts with $Q_R/Q_{RD}^*\,>\,1.2$.  This is responsible for the lack of data in the low-mass regime for $v_i\,=\,40$\,km\,s$^{-1}$ in Figure~\ref{fig:rd_comp}.

As shown in the top panel of Figure~\ref{fig:rd_comp}, the $CMF_{lr}$ vs. $m_{lr}$ plots are surprisingly similar for the \citetalias{Dou2024} and \citetalias{Reinhardt2022} formulations, despite the differences between their functional forms and $Q^*_{RD}$ formulae.  In general, it appears that, for regions in which data is available, \citetalias{Dou2024} tends to produce bodies with lower $CMF_{lr}$ values than \citetalias{Reinhardt2022}, though the decrease in CMF is typically less than 10\%. 

Another difference between these two formulations is in the size of targets/impactors necessary to produce remnants of a certain mass, as shown in the bottom panel of Figure~\ref{fig:rd_comp}.  It appears that \citetalias{Dou2024} requires larger impactors than \citetalias{Reinhardt2022} to produce resulting planetsimals of similar masses.  However, we as we are not considering a precise particle mass distribution in our calculations, this difference is unlikely to influence our later results.

\citetalias{Dou2024} also included prescriptions for oblique impact angles as opposed to \citetalias{Reinhardt2022}, which assumes that every collision is head-on.  High-velocity oblique impacts result in ``hit-and-run'' collisions, in which the two planetesimals effectively graze past each other.  They found that equal-mass collisions transitioned into the ``hit-and-run'' regime at the following velocity:

\begin{equation}
    v_\mathrm{HnR} = v_\mathrm{esc} \bigg( (-0.3 \times \mathrm{log}_{10} m_t  +2.8 ) \times (1-b)^{3.6} + 1.2 \bigg),
\end{equation}

\noindent where $b$ is the impact parameter.  If two objects were colliding with $0.2<b<0.7$ and $v_i>v_\mathrm{HnR}$, they defined separate mass and $CMF_{lr}$ relations for the resulting planetesimals.  Meanwhile, if the collision was slower than the hit-and-run velocity, they treated the results as perfect merging.  Finally, if $b<=0.2$, they found that they could approximate the impact as a head-on collision.  We once again will not enumerate the equations, but recommend reading \citetalias{Dou2024} for their exact functional forms and fit parameters. 

\begin{figure*}
    \centering
    \includegraphics[width=1\linewidth]{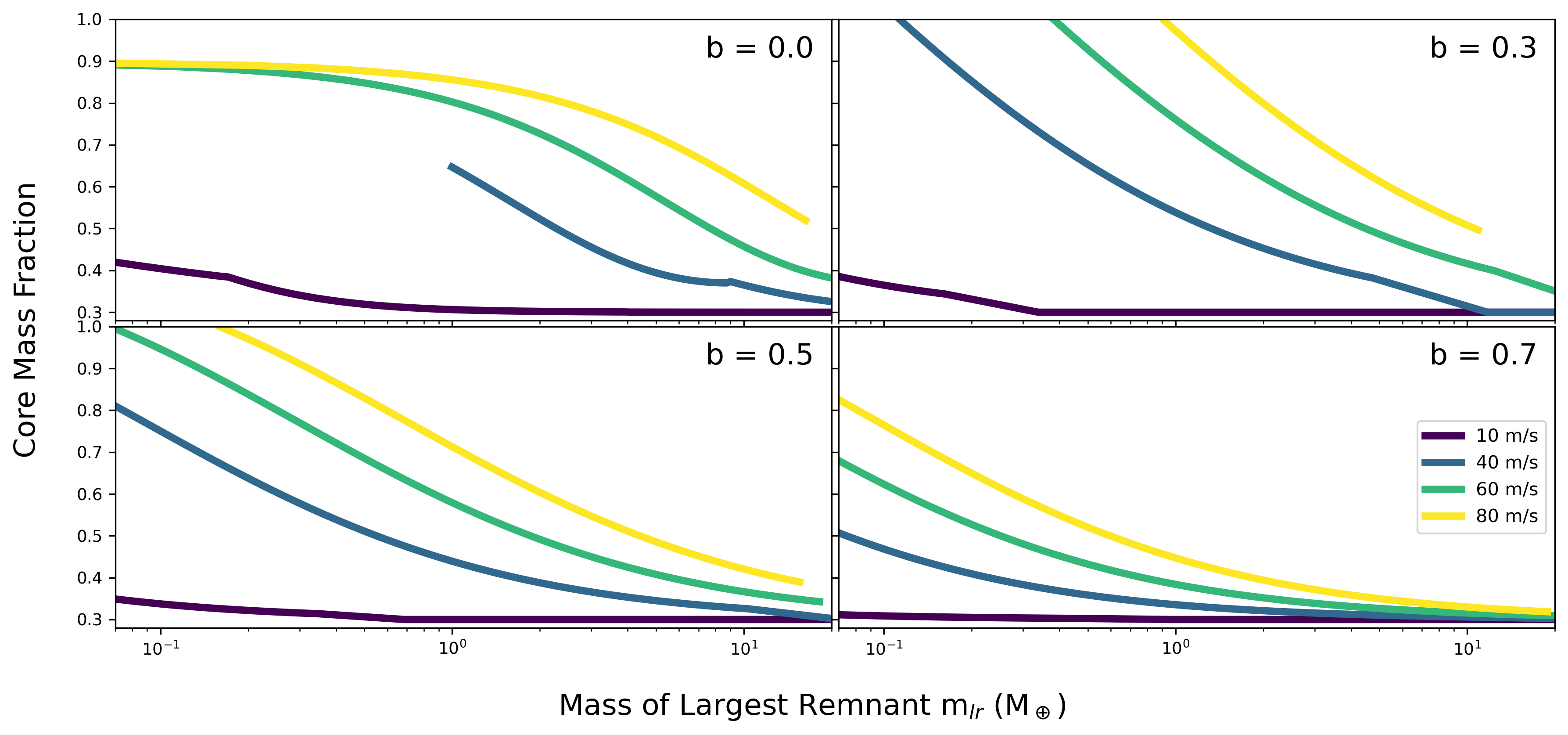}
    \caption{The core mass fraction as a function of $m_{lr}$ for various impact parameters $b$ using the \citetalias{Dou2024} formulation.  We utilize the same impact velocities as Figure~\ref{fig:rd_comp} for comparison.}
    \label{fig:cmf_mlr_obl}
\end{figure*}

We show the resulting $m_{lr}$ vs. $CMF_{lr}$ curves for oblique impacts in Figure~\ref{fig:cmf_mlr_obl}.  In general, we see that low impact parameters can result in an increase in $CMF_{lr}$ compared to head-on collisions, but high-$b$ impacts (such as $b\,\approx\,0.7$) tend to result in lower-$CMF_{lr}$ objects than the head-on case.  Given the dramatic effect that $b$ seems to have on the resulting planet $CMF_{lr}$, it is therefore necessary to track $b$ when simulating $v_\mathrm{imp}$ values in Section~\ref{ssec:coll} in order to get a more realistic impression of the ability of collisions to strip a planet's mantle.

\subsection{Collision Velocities}
\label{ssec:coll}

At this point, it is necessary to consider what factors influence the velocity of a planetesimal impact.  A planet's orbital velocity (assuming a circular orbit) is:

\begin{equation}
    v_\mathrm{circ} = \frac{2 \pi a}{P} = 213 \bigg(\frac{m_\star}{1\,M_\odot} \bigg)^{1/3} \bigg( \frac{P}{1\,\mathrm{d}}\bigg)^{-1/3}~\mathrm{km\,s}^{-1}.
\end{equation}

The orbital velocity increases as the host star mass increases and the orbital period decreases.  Therefore, we may expect to see higher collision velocities on short-period orbits around more massive stars, resulting in planets with more extreme $CMF_{lr}$ values.  However, it is unrealistic to assume that $v_\mathrm{imp}\,=\,v_\mathrm{circ}$, as most planetesimals will be orbiting in the same direction in the disk, and their orbits will not intersect unless they have some nonzero eccentricity or inclination term. As is stated in \cite{Lissauer1993} and derived in Appendix~\ref{app:vimp}, the impact velocity is proportional to both the inclinations and eccentricities of the planetesimals.  It is therefore important to consider what constitutes ``reasonable'' eccentricity and inclination distributions before proceeding any further.

\subsubsection{Eccentricity}
\label{sssec:ecc}

\cite{Stevenson2025} provided a very straightforward fit that we can use to describe the initial eccentricity distribution of the planetesimals.  However, as \cite{Stevenson2025} is a study of larger planets that primarily orbit older stars, their results may not accurately capture the eccentricity distribution of planetesimals in a young system.  Older systems have had more time to tidally circularize and may have already ejected or destroyed a large portion of any dynamically-hot population, so it is possible that this formulation is an underestimate of the eccentricities of planetesimals in young disks.

However, it is difficult to define a useful, one-size-fits-all eccentricity distribution for all planetesimals.  When performing simulations of planetesimals in an early disk, \citet{Agnor1999} found that the eccentricity distribution of planet-forming planetesimals evolves as a function of both time and planetesimal mass, where smaller bodies were more capable of achieving higher eccentricities, and that the typical eccentricity of these objects increased with time.  This is because the orbital elements of smaller bodies tend to be more dramatically affected by scattering events than larger bodies, imparting heightened eccentricities over time.  After running their simulations for 0.2\,Gyr, they found that the time-averaged eccentricity of planetesimals around 1\,$M_\oplus$ had ballooned to around 0.1, while the eccentricities of planetesimals below 0.1\,$M_\oplus$ was higher than 0.2.  As the \cite{Stevenson2025} eccentricity distribution has a mean around $e\,=\,0.16$, it may not necessarily be a terrible representation of the eccentricities of planets in the disk, though it neglects the relationship that eccentricity has with planetesimal mass.  Given the ambiguities regarding the exact time of planet formation, as well as the precise mass distribution of planetesimals in the early disk, we are hesitant to attempt to adopt a complex mass-and-time dependent function to describe the eccentricity of planets in our base model.

We may also take some inspiration from the eccentricity distribution of the asteroids in our own solar system, which may be representative of the planetesimal population in other systems.  \cite{Liu2023} analyzed over 400,000 main belt asteroids and found that the eccentricities were generally consistent with Rayleigh distributions with scale parameters of around 0.1, though inner-belt asteroids tended to deviate significantly from this distribution.  This Rayleigh distribution has a modal eccentricity of around 0.1 (higher than that of the Gamma distribution used otherwise) but results in the formation of extremely few planetesimals with eccentricities above 0.3, resulting in a lower mean eccentricity value.  However, once again, the generalizability of these observations are suspect, as they come from observations of old solar-system rocky bodies that, by definition, are on orbits stable enough to persist in the solar system for billions of years.  In addition, the orbits of the asteroid belt are likely heavily influenced by the gravity of the solar system planets (such as Jupiter), which may also influence their eccentricity.

Overall, when considering these sources, it appears that they primarily agree in terms of the shape and median of the eccentricity distribution (at least within a factor of two), with the majority of exoplanets possessing an eccentricity of 0.2 or less and then a small number of high-eccentricity objects.  It therefore appears that the bottom-heavy \cite{Stevenson2025} eccentricity distribution is a reasonable choice for our model if we assume a relatively dynamically quiet disk.

However, it is possible that, given its youth, the initial protoplanetary disk is very dynamically hot.  There are many scenarios that could impart enhanced eccentricities upon planetesimals in the inner disk, such as the orbital migration of nearby giant planets \citep{Carter2015}.  We therefore also consider a dynamically-hot planet formation scenario, where the planetesimals have heightened eccentricities, to study the impact of our initial eccentricity distribution on our results. Inspired by the eccentricity distributions shown in the figures of \cite{Carter2015} and \cite{Walsh2011} of planetesimals during phases of giant evolutionary phases, we will adopt a uniform eccentricity distribution between 0 and 1 as the ``dynamically hot'' scenario.  While this is likely an extreme case, it will at least provide us with an upper limit on the anticipated impact velocity distribution.

\subsubsection{Inclination}
\label{sssec:inc}

Planetesimal inclinations can also influence a planet's relative velocity, though we anticipate that our choice of planetesimal inclinations will have a weaker influence on our results than the planetesimal eccentricities, as the majority of collisions tend to occur at low inclination values (where the planetesimals' orbits are more likely to intersect).  However, it is still worthwhile to consider the reasonable limits of an inclination distribution in our calculations, as collisions can still occur at arbitrarily high planetesimal inclinations.

We can consider observations of our own solar system when trying to set a prior for the inclination distribution of planetesimals.  \cite{Brown2001} found that the Kuiper Belt is best-fit by a two-population model, one of objects with low inclinations of around 3$^\circ$ and a more dynamically-hot population with inclinations of around 15$^\circ$.  However, the solar system is a mature system, and therefore may not accurately represent the dynamical state of a young, forming system.

Given the brightness of the disks of young stars, we do have an opportunity to directly study the inclinations of particles in potential planet-forming regions.  $\beta$\,Pictoris is a young \citep[23\,Myr,][]{} system with a nearly edge-on disk, giving us an excellent opportunity to explore the vertical distribution of dust in a forming system.  The $\beta$\,Pictoris system appears to have two populations of planetesimals, with a dynamically-cold population with a root-mean-square $i$ value of around 1$^\circ$ and a dynamically-hot population with a root-mean-square $i$ of 9$^\circ$ \citep{Matra2019}.   These populations are also slightly misaligned relative to $\beta$\,Pictoris\,b's orbital plane, and \cite{Matra2019} concludes that the dynamically-excited planetesimals may be stirred up by perturbations from an additional planet in the system.  

The similarities between the $\beta$\,Pictoris system and the solar system seem to indicate that planetesimals are frequently organized into multiple populations, with dynamically-cold populations ($i\approx 0^\circ$) likely representing some natal distribution and dynamically-hotter ($i\approx 10-20^\circ$) populations representative of stirring by some additional body in the system.  Given the sensitivity of inclinations to the precise evolutionary and planet formation history of a system, it is difficult to generalize these findings to systems in general.  However, as a planetesimal inclination distribution of $i=\mathcal{U}(0^\circ, 20^\circ)$ seems to capture the majority of observed objects in $\beta$\,Pictoris and the solar system, we conclude that it is likely an acceptable simplification for our subsequent calculations.

\subsubsection{Calculating $v_\mathrm{imp}$}
\label{sssec:calc_vimp}

We estimate $v_\mathrm{imp}$ using the following method:

\begin{enumerate}
    \item Fix an orbital period and mass for the target body, assuming the host star has a mass of 1\,$M_\odot$.  Generate the target's orbital parameters following these steps:
    \begin{itemize}
        \item If we are assuming a dymamically cold scenario,, the eccentricity $e_t$ is drawn from a Gamma distribution with shape and scale parameters consistent with the distribution of super-Earths from \cite{Stevenson2025}.  We use the relationship for hot super-Earths if $P_t\,\leq\,10$d and the relationship for warm-super Earths if $P_t\,>\,10$d.  In the dynamically hot scenario, we allowed $e_t=\mathcal{U}(0,1)$.  Our reasoning behind these eccentricity distributions is described in Section~\ref{sssec:ecc}.
        \item The semi-major axis $a_t$ is calculated based on the fixed orbital period and host star mass.
        \item The argument of periastron $\omega_t$ is set equal to zero- we are interested in the relative motion between the target and impactor and not their absolute locations, so this is an acceptable simplification.
        \item The longitude of ascending node $\Omega_t$ is set to zero for similar reasons as $\omega_t$.
        \item The inclination $i_t$ is also set to zero, such that the inclination of the impactor $i_i$ is effectively the mutual inclination between the two objects.
    \end{itemize}
    \item Generate a potential impactor planetesimal.  The orbital elements are calculated as follows:
    \begin{itemize}
        \item The eccentricity $e_i$ is generated the same way as it is for the target body.
        \item $a_i$ is selected from a random uniform distribution such that it falls between 0 and 10\,AU.  While it is possible for there to be impactors that come from further out in the disk, we are only considering impacts with targets with semi-major axes less than 1 AU, so impacts with objects at $>10$\,AU are unlikely to be very common.
        \item The argument of periastron $\omega_i$ is $\mathcal{U}(0,\,2\pi)$.
        \item The longitude of ascending node $\Omega_i$ is $\mathcal{U}(0,\,2\pi)$.
        \item The inclination $i$ is drawn from a random uniform distribution between 0 and 20$^\circ$.  As we define the inclination of the target body as being at 0$^\circ$, this is the mutual inclination of the two bodies.  We discussed the selection of our inclination prior in Section~\ref{sssec:inc}.
        \item The mean anomaly $M_0$ at time $t_0$ is $\mathcal{U}(0,\,2\pi)$, and $t_0$ is $\mathcal{U}(0,\,P_i)$.
        \item The impactor mass $m_i$ is equal to that of the base planetesimal $m_t$.
    \end{itemize}.  
    \item Before the (computationally expensive) step of calculating the plantesimals' positions as a function of time, we checked to see if the impactor's orbit has a chance of intersecting the target's orbit by ensuring that the apastron of the impactor is greater than the periastron of the target and the periastron of the impactor is less than the apastron of target.  We also accounted for the possibility that the planets could interact even if their orbits do not directly intersect via gravitational focusing.  Explicitly, we rejected all orbits with the following criteria:
    \begin{equation}
    a_i< \frac{a_t(1-e_t) - 10(R_i+R_t)}{(1+e_i)};~~ a_i>\frac{a_t(1+e_t)+10(R_i+R_t)}{(1-e_i)}
    \end{equation}
    \noindent as we assume that gravitational focusing will typically not inflate the collisional cross-section by more than a factor of ten in most circumstances.
    \item Calculate the position and velocity of the target body as a function of time for 100 orbits using the Keplerian equation solver implemented in \texttt{radvel} \citep{Fulton2017, Fulton2018}, with a timestep equal to 1/100 of one orbital period of the target body.  We found that increasing the number of modeled orbits did not seem to dramatically affect our final results.
    \item Calculate the position and velocity of the impactor at the same time-steps as those of the target planet. After identifying the epoch of closest approach, we simulated 100 evenly-sampled times centered on this epoch.  We iteratively repeated this process until the timestep was such that the target planet did not move more than 1\,R$_\oplus$ between steps.
    \item We assume that the two planets collided if they passed within $R_\mathrm{coll}=(R_p+R_i)\sqrt{1+v_\mathrm{esc}^2/v_\mathrm{rel}^2}$ of one another at closest approach, with the second term accounting for the effects of gravitational focusing.  $v_\mathrm{esc}$ is the mutual escape velocity of the two bodies and $v_\mathrm{rel}$ is the relative Keplerian velocity between them.  We calculated the planet radii and escape velocities assuming that the planetesimals initially had Earth-like CMF values of $0.3$.
    \item We then calculated the X, Y, and Z velocities of each of the two bodies at collision, assuming that $v_\mathrm{coll} = \sqrt{v_\mathrm{rel}^2 + v_\mathrm{esc}^2}$ as is described in \cite{Moorhead2020}.
    \item We then calculated the impact parameter $b$ of the collision.  We defined $b$ as the sine of the angle between the impact velocity vector and the vector that connects the center of the two planetesimals when they are $R_\mathrm{coll}$ apart.
    \item We repeated steps i)- vii) for $10^9$ impactor-target pairs in order to calculate the distribution of impact velocities given the orbital period and target mass, which we could then use to find a distribution of $CMF_{lr}$ and $m_{lr}$ as a function of the these variables.
    \item We repeated steps i)-viii) for a grid of planet periods and masses, with $p_t$ = (0.3, 1, 3, 10, 30, 100, 300) days and $m_t$ = (0.1, 1, 10)\,$M_\oplus$.  
\end{enumerate}

Our simulations are not useful for calculating the absolute likelihood of collisions occurring, as we do not rigorously consider the mass distribution of planetesimals in the disk, and assume all collisions occur when $m_i=m_t$.  This is acceptable, as we are primarily interested in studying the distribution of $v_\mathrm{imp}$ as opposed to the absolute collision probability.  An extension of this work that includes a more realistic distribution of particles and particle masses as a function of semi-major axis would be straightforward (but time-consuming) to implement in the future. 

\begin{figure*}
    \centering
    \includegraphics[width=1\linewidth]{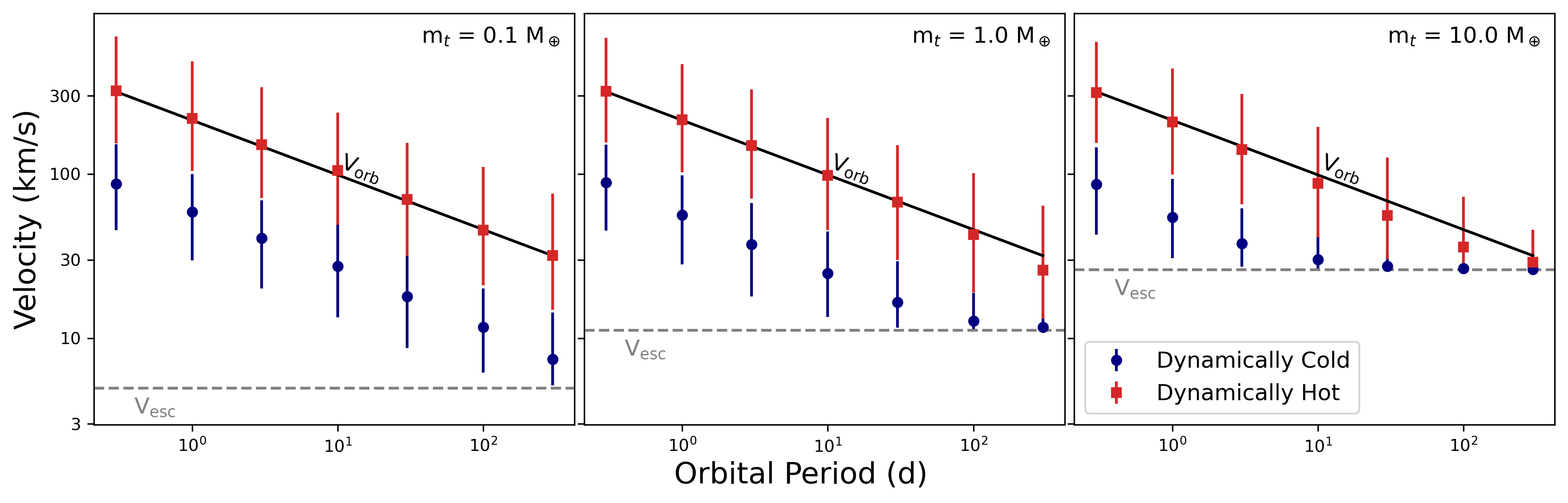}
    \caption{Impact velocity as a function of target orbital period in the dynamically cold (blue) and hot (red) scenarios for a variety of different target masses.  Each panel includes the orbital velocity as a function of period (solid black line) and the escape velocity associated with the target's mass (dashed gray line).}
    \label{fig:vimp_cold_hot}
\end{figure*}

Our results for the $m_\star\,=\,1\,M_\odot$ case (in both the dynamically cold and dynamically hot scenarios) are shown in gray in Figure~\ref{fig:vimp_cold_hot}.  In general, we see that, at short orbital periods, the typical impact velocity decreases sharply as orbital period increases, and at a similar rate as the orbital velocity itself (as in, $v_\mathrm{imp}\,\propto\,P^{-1/3}$), with the median $v_\mathrm{imp}$ at a given orbital period being about four times lower than the orbital velocity in the dynamically cold scenario, and comparable to the orbital velocity in the dynamically hot scenario.  The curve flattens around when $v_\mathrm{imp} \,=\,v_\mathrm{esc}$, as the impact velocity cannot drop below this value.  This may be useful as a general `rule of thumb' for estimating the typical impact velocities of planetesimals given their orbital parameters.

\begin{figure*}
    \centering
    \includegraphics[width=1\linewidth]{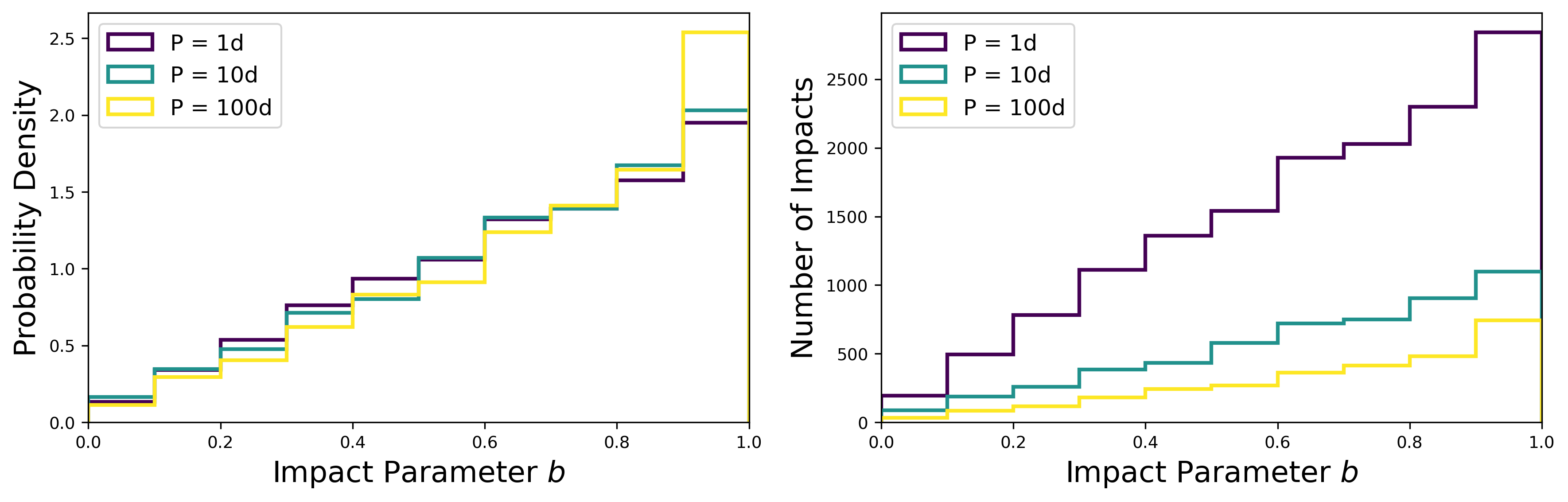}
    \caption{The probability distribution of the impact parameter $b$ from our impact calculations.  This figure specifically showcases the $b$ distribution for the $m_t=1\,M_\oplus$ planet in the dynamically cold case, though we note that the masses and eccentricity distribution of the planets had little influence on the overall $b$ distribution.  \textit{Left}.  A normalized histogram of $b$ for several different periods.  \textit{Right}.  The same histogram, but unnormalized to highlight the difference in number of impacts between different runs.}
    \label{fig:b_hist}
\end{figure*}

In Figure~\ref{fig:b_hist}, we show the distribution of impact parameters for our results for $m_p\,=\,1$\,$M_\oplus$ in the dynamically-cold scenario, though we note that the planet mass and starting eccentricity distribution of the disk did not have a strong influence on the impact parameter distribution.  Overall, we see that the majority of impacts occur at large impact parameters, with less than 10\% occurring at impact parameters below $b\,=\,0.2$.  As \citetalias{Dou2024} found that collisions with impact parameters above about $0.2$ required a different fit prescription than more head-on impacts, their formulation may be more useful for describing these collisions than that from \citetalias{Reinhardt2022}, which assumed head-on collisions (with $b\,\approx\,0$).

In addition, we note that the distribution of $b$ appears to be insensitive to the planet's orbital period, beyond the fact that there are far fewer collisions overall as we study longer and longer orbital periods.  This is logical given the small size of the planet compared to the size of its orbit.

\subsection{Connecting Collision Simulation Results}

We connected the results from our impact velocity calculations in Section~\ref{ssec:coll} to the models from both \citetalias{Dou2024} and \citetalias{Reinhardt2022} to estimate the maximum $CMF_{lr}$ of a planet as a function of the target mass, orbital period, and host star mass.  For the purposes of comparison, we assumed $\gamma = 1$ in these calculations.  A more optimal value of $\gamma$ would result in a higher $CMF_{lr}$ for a fixed mass and planet period in the \citetalias{Reinhardt2022} results, but this is acceptable as the difference is typically small in magnitude (as discussed in Section~\ref{ssec:r22}).

\begin{figure*}
    \centering
    \includegraphics[width=1\linewidth]{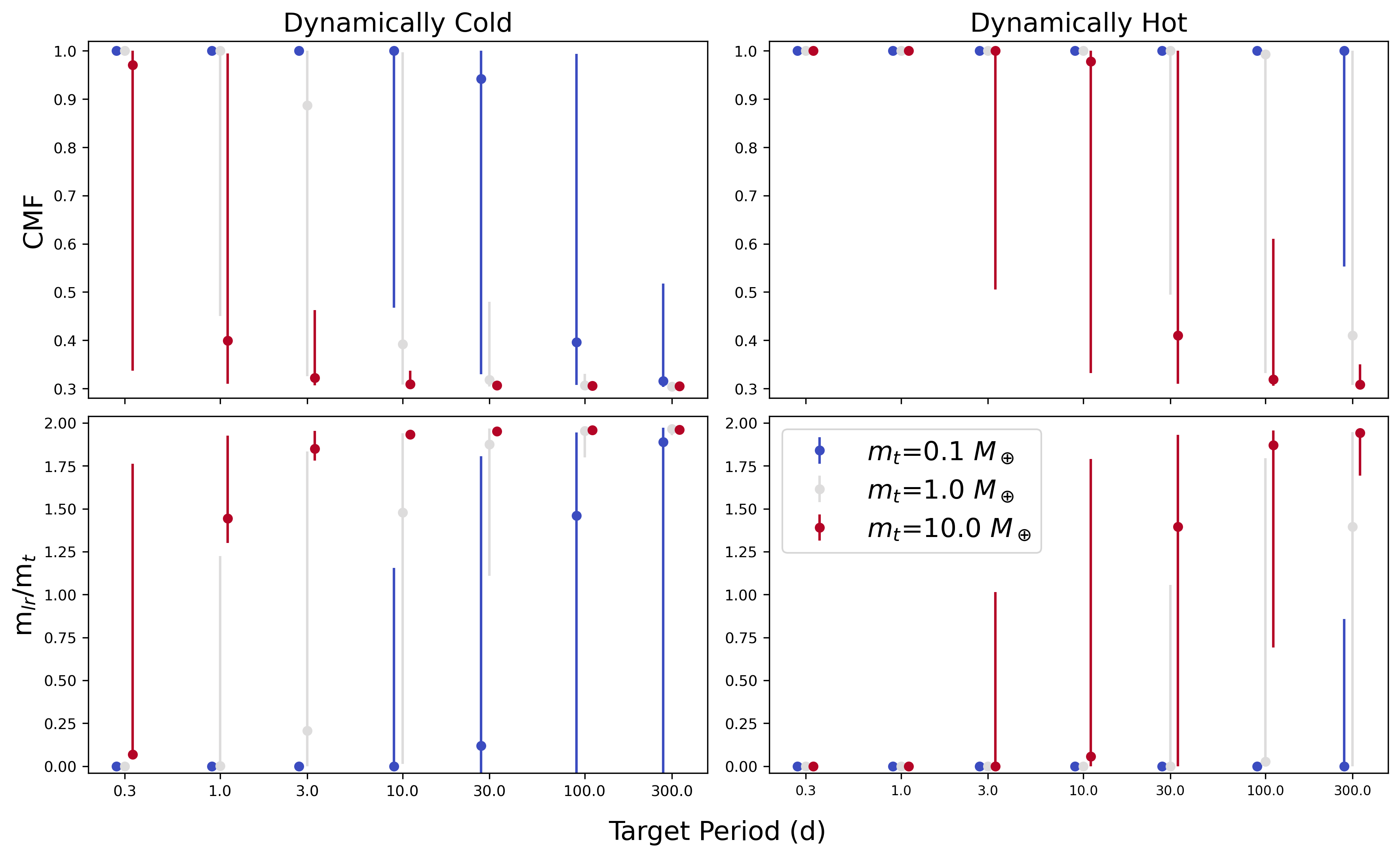}
    \caption{The inferred CMF (top) and $m_{lr}$ (bottom) as a function of planetary orbital period and target mass for our simulated collisions using \citetalias{Reinhardt2022}.  The left panels show our results in the case of a dynamically-cold disk, and the right panels show the same results in a dynamically-hot disk. This particular figure assumes a host star mass of 1\,M$_\odot$.}
    \label{fig:results_r22}
\end{figure*}

\begin{figure*}
    \centering
    \includegraphics[width=1\linewidth]{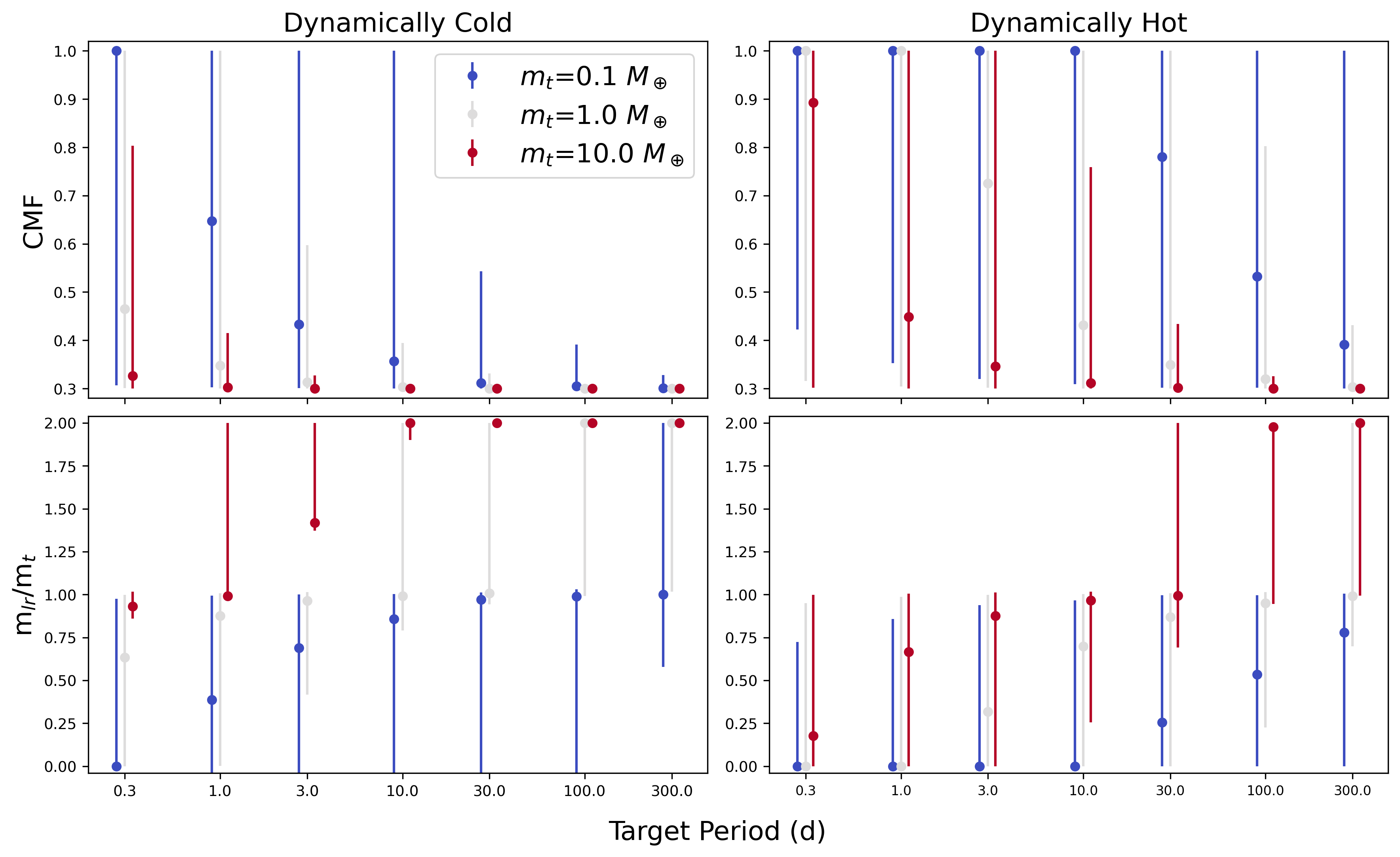}
    \caption{The same as Figure~\ref{fig:results_r22}, but with the collisions using the fit results from the \citetalias{Dou2024} simulations.}
    \label{fig:results_d24}
\end{figure*}

For each ``collision'' produced in Section~\ref{ssec:coll}, we calculated $CMF_{lr}$ and $m_{lr}$ using our calculated $b$, $v_\mathrm{imp}$, and simulated $m_t$ values. Our results in both the dynamically-cold and dynamically-hot scenarios are shown in Figures~\ref{fig:results_r22} and \ref{fig:results_d24}. We note that, given the different mass ranges for the two models, with $m_t\,=\,$\,1--20\,$M_\oplus$ for \citetalias{Reinhardt2022} and $m_t\,=\,$\,0.06--20\,$M_\oplus$ for \citetalias{Dou2024}, we can only study sub-Earth collisions using the \citetalias{Dou2024} models.  In addition, as the functions in \citetalias{Dou2024} can create nonphysical values for $m_{lr}$ and $CMF_{lr}$ at very high velocities, we placed an upper bound of $CMF_{lr}=1$.

These figures reveal several simple trends.  Firstly, low-velocity collisions at larger orbital periods tend to result in perfect mergers, resulting in a final planet with a mass equal to $m_t+m_i \,=\, 2 m_t$ and the same CMF as the two constituent bodies.  Secondly, it is difficult to produce very high-mass bodies with large CMF values, as the high-velocity collisions that could produce iron-rich bodies tend to cause significant fragmentation (such that $m_{lr}$ is very low).

In general, as expected, we found that the \citetalias{Dou2024} simulations produced lower CMF values than those from \citetalias{Reinhardt2022}.  This is likely due to the fact that \citetalias{Dou2024} contains a more realistic prescription for off-axis impacts, while the \citetalias{Reinhardt2022} simulations assumed that all impacts are head-on.  Looking at the $b$ distribution of impacts that we modeled Figure~\ref{fig:b_hist}, we see that $<10\%$ of all collisions occur at impact parameters $<0.2$, meaning that the \citetalias{Dou2024} prescription is more useful for describing the results of the majority of the planetesimal impacts in our study.  In addition, given the relatively high minimum target mass ($1\,$M$_\oplus$) considered in the \citetalias{Reinhardt2022} simulations, their results are poor at describing sub-Earths, as the majority of impacts that produce sub-Earths require target masses outside of their range of consideration. Meanwhile, with a minimum $m_t$ of 0.06\,M$_\oplus$, \citetalias{Dou2024} has target masses that are substantially better at exploring the sub-Earth regime, only struggling to reproduce the effects of extremely high-energy impacts (where their fit equations result in a CMF$\,>\,1$).  

It is necessary to consider some of the \citetalias{Dou2024} model caveats.  Firstly, \citetalias{Dou2024} states that their simulations produced results too small to track at low target masses and high velocities.  \citetalias{Dou2024} states that these fragmentation collisions occurred in head-on ($b<0.2$) collisions with $m_t$\,<2\,M$_\oplus$ and $Q_R/Q^*_{RD}$\,>\,1.2.  Examining our results, we found that these collisions made up less than 2\% of the total collisions modeled in both the dynamically-hot and dynamically-cold scenarios, and are therefore unlikely to contribute substantially to our results.  In addition, the fits from \citetalias{Dou2024} are to data with $b$ varying from 0 -- 0.7, and therefore the prescription might be less accurate for bodies with higher impact parameters, which include about half of the bodies considered (see Figure~\ref{fig:b_hist}).  However, we have no reason to expect that planetary impacts at higher impact parameters would behave substantially differently than those at lower impact parameters.

As anticipated, we find that lower-mass planets on shorter-period orbits are capable of achieving higher $CMF_{lr}$ values.  In general, it appears that it is difficult to form super-Earths with Mercury-like $CMF_{lr}$ values unless they have extremely short orbital periods.  It is far easier to form sub-Earths with high CMFs.  Sub-Earths can retain Mercury-like CMF values out to orbital periods of tens of days, which follows the wisdom that it is easier to strip away the mantles of low-mass planets.  In addition, given the relationship between orbital velocity and host star mass, the host star mass can also influence the CMFs of its planets, with more massive stars hosting higher-CMF planets at similar orbital distances than planets around smaller stars.

\subsection{The Influence of Eccentricity and Inclination}
\label{ssec:inc_ecc}

When simulating the planet collisions, we performed calculations assuming both a dynamically cold (with an eccentricity distribution informed by observed mature systems) and a dynamically hot (with a uniform distribution in eccentricity) system.  As shown in Figure~\ref{fig:vimp_cold_hot}, a dynamically hot system will have significantly higher impact velocities than a dynamically cold system.  The left and right panels of Figures~\ref{fig:results_r22} and \ref{fig:results_d24} compare the $m_{lr}$ and CMF values for the dynamically cold and hot cases.  It is clear that the higher impact velocities imparted in the dynamically-hot scenario result in collisions that form lower-mass bodies with higher CMFs.

\begin{figure}
    \centering
    \includegraphics[width=1\linewidth]{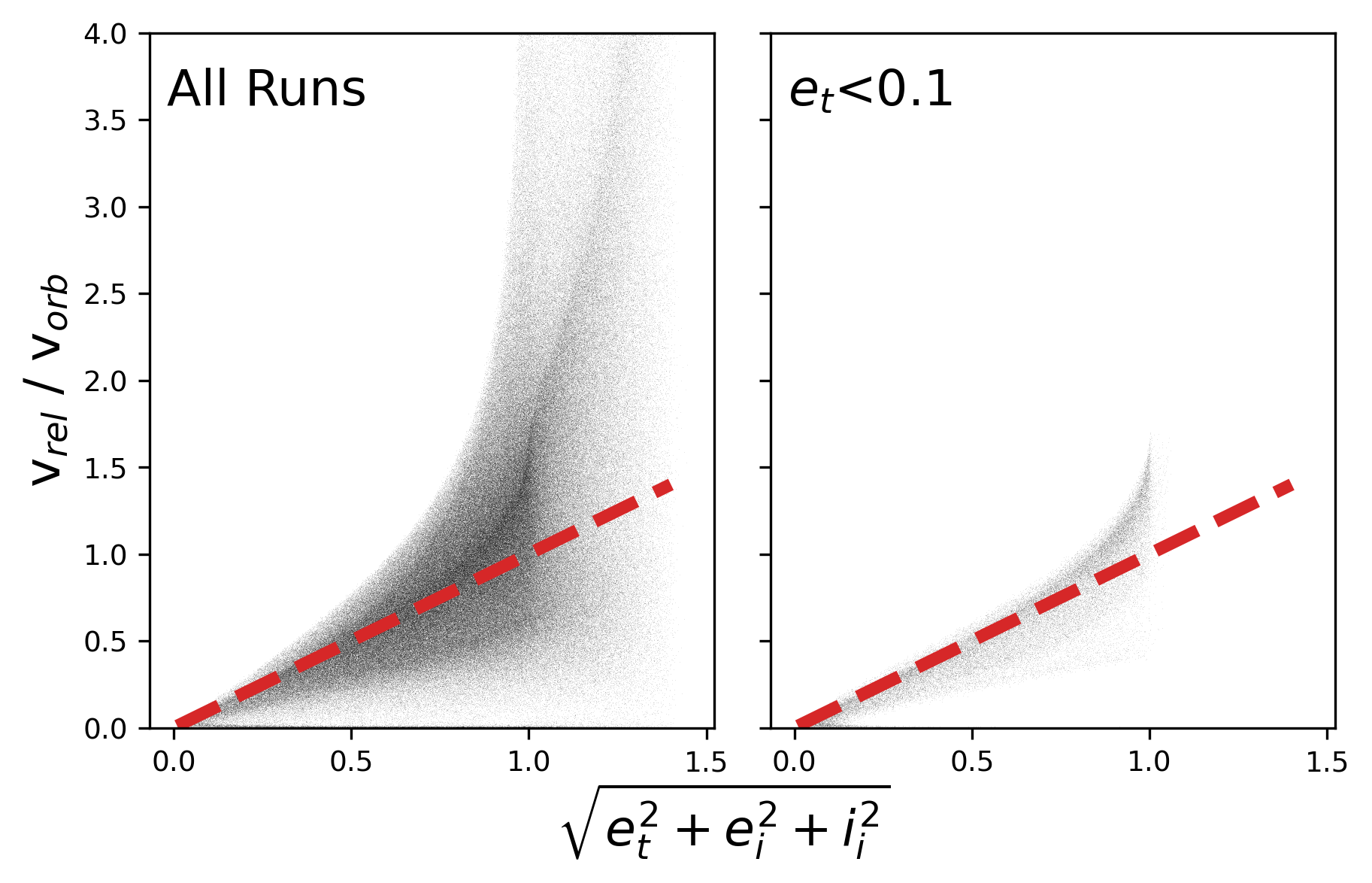}
    \caption{The normalized relative velocity ($v_\mathrm{rel}/v_\mathrm{orb}$) as a function of $\sqrt{e_t^2+e_i^2+i_i^2}$ for our impacts, where $i_i$ is in radians. The impact velocity is straightforward to calculate from $v_\mathrm{rel}$ by adding $v_\mathrm{esc}$ in quadrature. The points are colored according to the number of collisions that occurred with those parameters.  We also include a red dashed line at x=y to visualize where our model ceases to accurately describe the data.}
    \label{fig:vimp_sini_ecc}
\end{figure}

Given the obvious influence of the eccentricity distribution on our results, we attempted to construct a function that relates the impact velocity and the planetesimals' eccentricities.  As shown in Figure~\ref{fig:vimp_sini_ecc}, the observed impact velocity is a strong function of both the inclination and eccentricity of the impacting body.  Inspired by the $v_{imp}\approx v_\mathrm{kep}\sqrt{e^2+i^2}$ formula from \cite{Lissauer1993}, we found that $v_\mathrm{imp}/v_\mathrm{orb}\approx\sqrt{e_t^2 +e_i^2+i_i^2}$ when $v_\mathrm{imp}/v_\mathrm{orb}<0.4$ (and $i_i$ is in radians), though there is some scatter to this relationship.  We discuss the potential source of this scaling relationship in Section~\ref{ssec:analytic}.

Therefore, overall, we observe that both the eccentricity and inclination of the impacting body are directly related to $v_\mathrm{imp}$, with higher values of inclination and eccentricity resulting in higher-energy impacts.  This means that the resulting $v_\mathrm{imp}$ distribution of our simulations is strongly influenced by our assumptions regarding the eccentricities and inclinations of the impacting planetesimals. This explains the fact that we observed higher impact velocities in our dynamically-hot simulations than in our dynamically-cold simulations.  These changes can be quite dramatic- as shown in Figure~\ref{fig:vimp_sini_ecc}, doubling the eccentricity of a planetesimal could more than double its impact velocity, which would have significant implications regarding its ability to strip the target planet's mantle. 

\begin{figure*}
    \centering
    \includegraphics[width=1\linewidth]{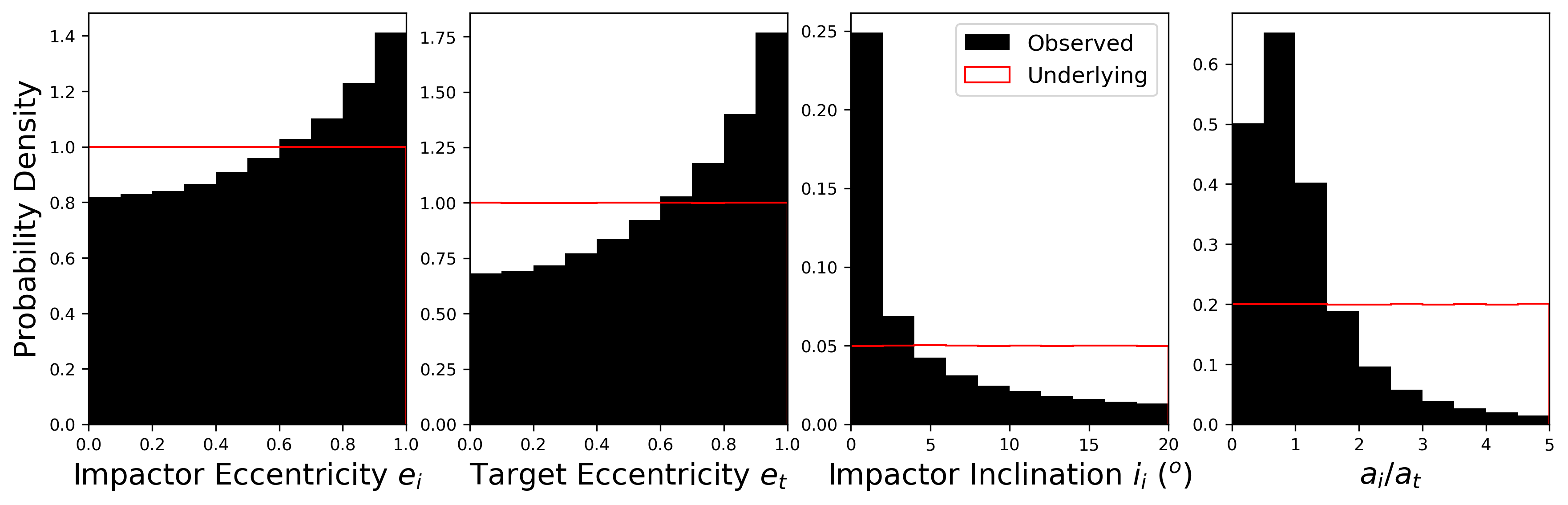}
    \caption{The histograms of the eccentricities, inclinations, and semi-major axis ratios of the planetesimals in the dynamically-hot simulation that impacted the target, in black.  We also include, as a red histogram, the underlying distributions of these parameters that we used to generate the planetesimal population.}
    \label{fig:nimpacts_ecc_inc.png}
\end{figure*}

We can also use these simulations to study how the planetesimal parameters influence their impact probability.  We do this by comparing the distribution of orbital parameters of the planets that underwent impacts to the underlying simulated parameter distributions.  Figure~\ref{fig:nimpacts_ecc_inc.png} shows the resulting distributions for $e_i$, $e_t$, $i_i$, and $a_i$ for our dynamically-hot simulations.

As shown in Figure~\ref{fig:nimpacts_ecc_inc.png}, planetesimals with high eccentricities are more likely to undergo collision. This is likely due to the fact that planetesimals with higher eccentricities are more likely to have planet-crossing orbits, resulting in more impacts.  As higher eccentricities are more likely to result in impacts, our modeled eccentricity distribution has a very strong influence on our final modeled $v_\mathrm{imp}$ distribution.

In addition, the mutual inclinations of the planetesimals has an influence on their impact probability.  While we simulated a uniform distribution of inclinations, we observed a large concentration of impacts at low inclinations, which levels off to a roughly uniform distribution of impact inclinations at higher inclinations (above about $i\approx10^\circ$).  This can be understood by considering the orbits of planets as two ellipses. If the mutual inclination is nonzero, the impactor crosses through the plane of the target body's orbit in exactly zero or two places.  However, if the mutual inclination between the two bodies is low, the nonzero radius of the two bodies can allow for them to collide even if the two orbits do not perfectly intersect.  However, at higher mutual inclinations,  the collisions typically only occur right as the planet passes through the midplane, and the number of these crossings are fully independent of the planet's inclination.  Therefore, at high mutual inclinations, the probability of impact is insensitive to the actual inclination value. This means that our results are not very sensitive to our modeled inclination distribution, as the majority of impacts occur at low inclinations.

Finally, it appears that the majority of our collisions occur when $a_i\approx a_t$, which is logical as these orbits are the most likely to intersect.  As $a_i$ gets very large, the probability of impacts dramatically decreases, which may be a side-effect of the finite length of our simulation runs.  However, we do note that when $a_i$ is very close to $a_t$ (differing by less than 1\%), there is a decrease in impact probability, likely related to the fact that planets on identical, low-eccentricity orbits are unlikely to intersect given their nearly identical periods.

\subsection{Analytic Estimates}
\label{ssec:analytic}

We can analytically estimate the velocity of a collision between a target body and an impactor as follows.  Firstly, as discussed in \cite{Schwarz2017} and reproduced in Appendix~\ref{app:pv}, the position $p$ and velocity $v$ of a body at a given time can be described as a function of the its semi-major axis ($a$), eccentricity ($e$), inclination ($i$), true anomaly ($f(t)$), argument of periastron ($\omega$), and longitude of ascending node ($\Omega$).  We can then determine $v_\mathrm{rel}$ with:

\begin{equation}
\label{eqn:vimp_def}
    v_\mathrm{rel} = \sqrt{\big(v_{x,t}-v_{x,i}\big)^2 +\big(v_{y,t}-v_{y,i}\big)^2 + \big(v_{z,t}-v_{z,i}\big)^2}.
\end{equation}

To include the influence of gravitational focusing, we would follow the formulation of \cite{Moorhead2020} and calculate

\begin{equation}
    v_\mathrm{imp} \approx \sqrt{v_\mathrm{rel}^2 + v_\mathrm{esc}^2}, 
\end{equation}

\noindent where $v_\mathrm{rel}$ is influenced by the orbital parameters of the bodies and $v_\mathrm{esc}$ (the mutual escape velocity of the planetesimals) is time-independent and is only a factor of the sizes and masses of the planetesimals.  Given the ease of calculating $v_\mathrm{esc}$, we will focus on calculating $v_\mathrm{rel}$ in this section and only include $v_\mathrm{esc}$ as a final step.

Before calculating $v_\mathrm{rel}$, however, we must determine whether or not the orbits intersect (as non-intersecting orbits make impacts impossible) and then the specific time $\tau_\mathrm{imp}$ at which the two bodies impact, which will allow us to find $f(\tau)$ and fully define all of the terms in $v_\mathrm{rel}$.  

To simplify our math, we will assume the target body has $i_t=\omega_t=\Omega_t=0$, in which case the impacting body's orbit is defined relative to it.  With this simplification, we find that the position of the target is:

\begin{equation}
    p_{x,t} (\tau) = a_t \bigg(1-e_t\,\mathrm{cos} \big(E_t(\tau)\big) \bigg) \mathrm{cos}\big(f_t(\tau)\big),
\end{equation}

\begin{equation}
    p_{y,t} (\tau) = a_t \bigg(1-e_t\,\mathrm{cos} \big(E_t(\tau)\big) \bigg) \mathrm{sin}\big(f_t(\tau)\big),
\end{equation}

\noindent and (as $i_t=0$):

\begin{equation}
    p_{z,t} (\tau) = 0,
\end{equation}

\noindent where $E_t(\tau)$ is the eccentric anomaly of the target body at time $\tau$.

Meanwhile, the velocity is of the target is:

\begin{equation}
\label{eqn:vxt}
    v_{x,t} (\tau) = -\frac{\sqrt{GM a_t}}{a_t \bigg(1-e_t \mathrm{cos}\big(E_t(\tau)\big)\bigg)} \mathrm{sin}\big(E_t(\tau)\big),
\end{equation}

\begin{equation}
\label{eqn:vyt}
    v_{y,t} (\tau) = \frac{\sqrt{GM a_t(1-e_t^2)}}{a_t \bigg(1-e_t \mathrm{cos}\big(E_t(\tau)\big)\bigg)} \mathrm{cos}\big(E_t(\tau)\big),
\end{equation}

\noindent and:

\begin{equation}
    v_{z,t}  (\tau)= 0.
\end{equation}

Moving forward, we will omit the $\tau$ dependence from these equations in order to reduce their visual complexity.  We note that $\tau$ moving forward will always be the time of impact.

Now, we must consider what conditions must hold for the two objects to collide.  If we assume that both of the bodies are infinitesimally small, then they will only collide if their orbits intersect.  This happens when $p_{x,t}=p_{x,i}$, $p_{y,t}=p_{y,i}$, and $p_{z,t}=p_{z,i}$.  As $p_{z,t}$ is equal to zero, this allows us to easily derive some constraints on the orbital elements of the impactor.

First, we can set $p_{z,i}=0$:

\begin{equation}
\begin{aligned}
     0=  a_i \bigg(1-e_i\,\mathrm{cos} \big(E_i\big) \bigg) \times \\\Bigg(\mathrm{cos\big(f_i\big)} \mathrm{sin}\big(\omega_i\big) \mathrm{sin}\big(i_i\big)  + \mathrm{sin\big(f_i\big)} \mathrm{cos}\big(\omega_i\big)\mathrm{sin}\big(i_i)\Bigg)
\end{aligned}
\end{equation}

This equation immediately has a trivial solution when sin$(i_i)$ = 0, which occurs when the impactor has the same inclination as the target, and they are both at $p_{z}=0$ at all times.

To find additional nontrivial solutions, we can simplify the equation:

\begin{equation}
    -\mathrm{cos\big(f_i\big)} \mathrm{sin}\big(\omega_i\big)  =\mathrm{sin\big(f_i\big)} \mathrm{cos}\big(\omega_i\big)
\end{equation}

\noindent and find:

\begin{equation}
\label{eqn:nu_i}
    f_i = -\omega_i + n\pi,
\end{equation}

\noindent where $n$ is an integer.  Interestingly, this allows us to dramatically simplify the equations for $p_{x,i}$ and $p_{y,i}$ at the time of impact to:

\begin{equation}
\label{eqn:pxi}
    p_{x,i} =  \pm a_i \bigg(1-e_i\,\mathrm{cos} \big(E_i\big) \bigg) \mathrm{cos}\big(\Omega_i\big)
\end{equation}

\noindent and:

\begin{equation}
\label{eqn:pyi}
    p_{y,i} =   \pm a_i \bigg(1-e_i\,\mathrm{cos} \big(E_i\big)\bigg) \mathrm{sin}\big(\Omega_i\big)
\end{equation}

\noindent where the positive solution is true when $n$ is even and the negative solution is true when $n$ is odd.

Now, the two planets will collide if $p_{x,i}=p_{x,t}$ and $p_{y,i}=p_{y,t}$:

\begin{equation}
\label{eqn:ai_at}
    \pm a_i \bigg(1-e_i\,\mathrm{cos} \big(E_i\big)\bigg) \mathrm{cos}\big(\Omega_i\big) = a_t \bigg(1-e_t\,\mathrm{cos} \big(E_t\big)\bigg) \mathrm{cos}\big(f_t\big)
\end{equation}

\noindent and:

\begin{equation}
   \pm  a_i \bigg(1-e_i\,\mathrm{cos} \big(E_i\big)\bigg) \mathrm{sin}\big(\Omega_i\big) = a_t \bigg(1-e_t\,\mathrm{cos} \big(E_t\big)\bigg) \mathrm{sin}\big(f_t\big).
\end{equation}

When $n$ is even, this straightforwardly simplifies to:

\begin{equation}
\label{eqn:nu_omega_even}
f_t = \Omega_i + 2 m\pi
\end{equation}

\noindent for some $m \in \mathbb{Z}$. 

However, when $n$ is odd, this instead simplifies to:

\begin{equation}
\label{eqn:nu_omega_odd}
f_t = \Omega_i + \pi + 2 m\pi.
\end{equation}

While we now know the value of $f_t$, the velocity of the target is a function of $E_t$, not $f_t$ (see Equations~\ref{eqn:vxt} and \ref{eqn:vyt}).  To calculate $E_t$, we can utilize the definition of the true anomaly:

\begin{equation}
\label{eqn:nu_def}
    f_t = 2\,\mathrm{tan}^{-1}\Bigg(\frac{\sqrt{1+e_t}}{\sqrt{1-e_t}\,} \mathrm{tan}\bigg(\frac{E_t}{2}\bigg)\Bigg),
\end{equation}

\noindent and find:

\begin{equation}
\label{eqn:Et_def}
    E_t = 2\,\mathrm{tan}^{-1}\Bigg(\frac{\sqrt{1-e_t}}{\sqrt{1+e_t}\,} \mathrm{tan}\bigg(\frac{f_t}{2}\bigg)\Bigg) 
\end{equation}

We can then insert Equation~\ref{eqn:nu_omega_odd} (when $n$ is odd) or Equation~\ref{eqn:nu_omega_even} into Equation~\ref{eqn:Et_def}:
$$
E_t = \begin{cases}
    2\,\mathrm{tan}^{-1}\Bigg(\frac{\sqrt{1-e_t}}{\sqrt{1+e_t}\,} \mathrm{tan}\bigg(\frac{\Omega_i }{2}\bigg)\Bigg) & \text{if } ~n~\mathrm{even}\\
        2\,\mathrm{tan}^{-1}\Bigg(\frac{\sqrt{1-e_t}}{\sqrt{1+e_t}\,} \mathrm{tan}\bigg(\frac{\Omega_i + \pi}{2}\bigg)\Bigg)& \text{if } ~n~\mathrm{odd}.
\end{cases}
$$

As tan($\pi + x$) = tan($x$), this is functionally equivalent to:

\begin{equation}
\label{eqn:Et_def_omega}
    E_t = 2\,\mathrm{tan}^{-1}\Bigg(\frac{\sqrt{1-e_t}}{\sqrt{1+e_t}\,} \mathrm{tan}\bigg(\frac{\Omega_i + n\pi}{2}\bigg)\Bigg).
\end{equation}

By substituting Equation~\ref{eqn:Et_def_omega} into Equations~\ref{eqn:vxt} and \ref{eqn:vyt}, we can calculate $v_{x,t}$ and $v_{y,t}$ at the time of impact as a function of the planets' orbital elements.  

To calculate the velocity of the impactor, we can utilize the fact that, as follows naturally from Equations~\ref{eqn:ai_at}, \ref{eqn:nu_omega_even}, and \ref{eqn:nu_omega_odd}:

\begin{equation}
\label{eqn:dt_di}
     a_i \bigg(1-e_i\,\mathrm{cos} \big(E_i\big)\bigg) = a_t \bigg(1-e_t\,\mathrm{cos} \big(E_t\big)\bigg).
\end{equation}

We can then solve this equation for $E_i$, finding:

\begin{equation}
E_i = \pm\mathrm{cos}^{-1}\bigg(\frac{a_i-a_t + a_t e_t\,\mathrm{cos} \big(E_t\big)}{e_i a_i}\bigg).
\end{equation}

The final constraint comes from the definition of the true anomaly (Equation~\ref{eqn:nu_def}) for the impacting body, combined with our understanding of $f_i$ from Equation~\ref{eqn:nu_i}:

\begin{equation}
\label{eqn:omega_condition}
    \omega_i = \mp 2\,\mathrm{tan}^{-1}\Bigg(\frac{\sqrt{1+e_i}}{\sqrt{1-e_i}\,} \mathrm{tan}\bigg(\frac{1}{2} \mathrm{cos}^{-1}\bigg(\frac{a_i-a_t + a_t e_t\,\mathrm{cos} \big(E_t\big)}{e_i a_i}\bigg) \bigg)\Bigg) + n\pi,
\end{equation}

\noindent where, as a reminder, $E_t$ is purely a function of $e_t$ and $\Omega_i$.  We note that this means that there are up to four values of $\omega_i$ at which the orbits can cross- the solutions for $\pm E_i$ and the solutions for an even and odd $n$.

We now have all the pieces we need to calculate $v_\mathrm{rel}$ (as described in Equation~\ref{eqn:vimp_def}) as a function of $a_t$, $a_i$, $e_t$, $e_i$, and $\Omega_i$.  As the equation is lengthy, we will not reproduce it fully in this text.  However, we can write out $v_\mathrm{rel}$ with the simplifying assumption that the target body has a circular orbit (that is, $e_t=0$):

\begin{equation}
\begin{aligned}
\label{eqn:impact_velocity}
    v_\mathrm{rel} = v_{orb,t} \Bigg(3 - \frac{a_t}{a_i} \mp \frac{2 }{\sqrt{a_i a_t} e_i}  \mathrm{cos}\big(i_i\big) \times \\ \bigg(  (a_i-a_t) \sqrt{1-e_i^2} \mathrm{cos} \big(\omega_i\big) \mp \sqrt{-(a_i-a_t)^2 + a_i^2 e_i^2} \mathrm{sin} \big(\omega_i\big)\bigg) \Bigg)^{1/2}
\end{aligned}
\end{equation}

\noindent with the value for $\omega$ defined in Equation~\ref{eqn:omega_condition}.   The first $\mp$ term corresponds to the value of $n$, with the ``minus'' in the case where $n$ is even and the ``plus'' in the case where $n$ is odd.  Meanwhile, the second $\mp$ term corresponds to the value of $E_i$, where the ``minus'' term corresponds to a positive $E_i$ while the ``plus'' term corresponds to a negative $E_i$.

This allows us to calculate the relative velocity given only the eccentricity, semi-major axis, and inclination of the target.  As $\cos(-\omega_i) = \cos(\omega_i)$ and $\sin(-\omega_i) = -\sin(\omega_i)$, $v_\mathrm{rel}$ is the same whether we select the negative or the positive value for $E_i$ in this case.  We therefore will omit the negative $E_i$ solution from our math in the $e_t=0$ case without a loss of generality.  The selection of $n$ being even or odd similarly has little impact on $v_\mathrm{rel}$ in the $e_t=0$ case.  While this expression is difficult to simplify further, we can simplify it if we assume that the eccentricity is low, forcing $a_i\approx a_t$.  Our resulting impact velocity formula, which we provide a derivation for in Appendix~\ref{app:vimp}, is identical to that in \cite{Lissauer1993}.  For low values of $e_i$ and $i_i$,

\begin{equation}
    v_\mathrm{rel} \approx v_{orb,t} \sqrt{e_i^2 + i_i^2 }.
\end{equation}

In general, this agrees with our results from our simulations (shown in Figure~\ref{fig:vimp_sini_ecc}), which show a correlation between these values for our simulations when $e_t$ is low.  Unfortunately, it is difficult to provide a similarly simple equation in the case where $e_t>0$, though $v_\mathrm{imp}$ is still straightforward to calculate analytically.

Using Equation~\ref{eqn:impact_velocity}, we can now compare our theoretical estimates with our results from the simulated collisions.  To do so, we analytically calculated the value of $v_\mathrm{imp}$ given an arbitrary set of target and impactor properties (primarily $a_t$, $a_i$, $i_i$, $e_i$, and $e_t$, similar to our simulations) using the same priors on $a_t, i_i, \Omega_i$, $e_i$, and $e_t$ as in our simulations (Section~\ref{ssec:coll}) and fixing $\omega_i$ to the value calculated in Equation~\ref{eqn:omega_condition}, randomly selecting which of the four valid solutions of $\omega_i$ to use for each run.  We note that our bounds on $a_i$ follow directly from Equation~\ref{eqn:dt_di}:

\begin{equation}
     a_i  = \frac{a_t \bigg(1-e_t\,\mathrm{cos} \big(E_t\big)\bigg)}{\bigg(1-e_i\,\mathrm{cos} \big(E_i\big)\bigg)}.
\end{equation}

By the definition of the trigonometric function, the value cos($E_i$) can only vary between -1 and 1.  Therefore, the limits on $a_i$ such that collisions can happen are: 

\begin{equation}
\label{eqn:ai_limits}
\frac{a_t \bigg(1-e_t \,\mathrm{cos} \big(E_t\big)\bigg)}{1+e_i} \leq a_i \leq \frac{a_t  \bigg(1-e_t\,\mathrm{cos} \big(E_t\big)\bigg)}{1-e_i}.
\end{equation}

These are comparable to the limits on $a_i$ that we used in our simulations, though we allowed for less restrictive limits in that case, as the planetesimals in our simulations are not infinitesimally small.

With these limits on $a_i$, we can estimate the maximum possible velocity that can be attained by an impact, when $i_i\approx90^\circ$ and $e_i\approx1$.  In this case,

\begin{equation}
\begin{aligned}
    v_{imp,max} = v_{orb,t} \Bigg( \frac{3+e_t}{1-e_t} - \frac{a_t}{a_i} \Bigg)^{1/2}.
\end{aligned}
\end{equation}

Substituting in the maximum possible value for $a_i$ (which is, in this case, infinite as $e_i$ approaches 1):

\begin{equation}
\begin{aligned}
    v_{imp,max} = v_{orb,t} \Bigg( \frac{3+e_t}{1-e_t}\Bigg)^{1/2}.
\end{aligned}
\end{equation}

When $e_t=0$, we find that $v_{imp,max} = \sqrt{3}v_{orb,t}$.  However, when $e_t$ is allowed to vary, $v_{imp,max}$ is infinite as $e_t$ approaches 1.  More practically, $v_{imp,max} = 2 v_{orb,t} (1-e_{t,max})^{-1/2}$, where $e_{t,max}$ is the maximum value of $e_t$ that we allow for in our simulations.

\begin{figure*}
    \centering
    \includegraphics[width=1\linewidth]{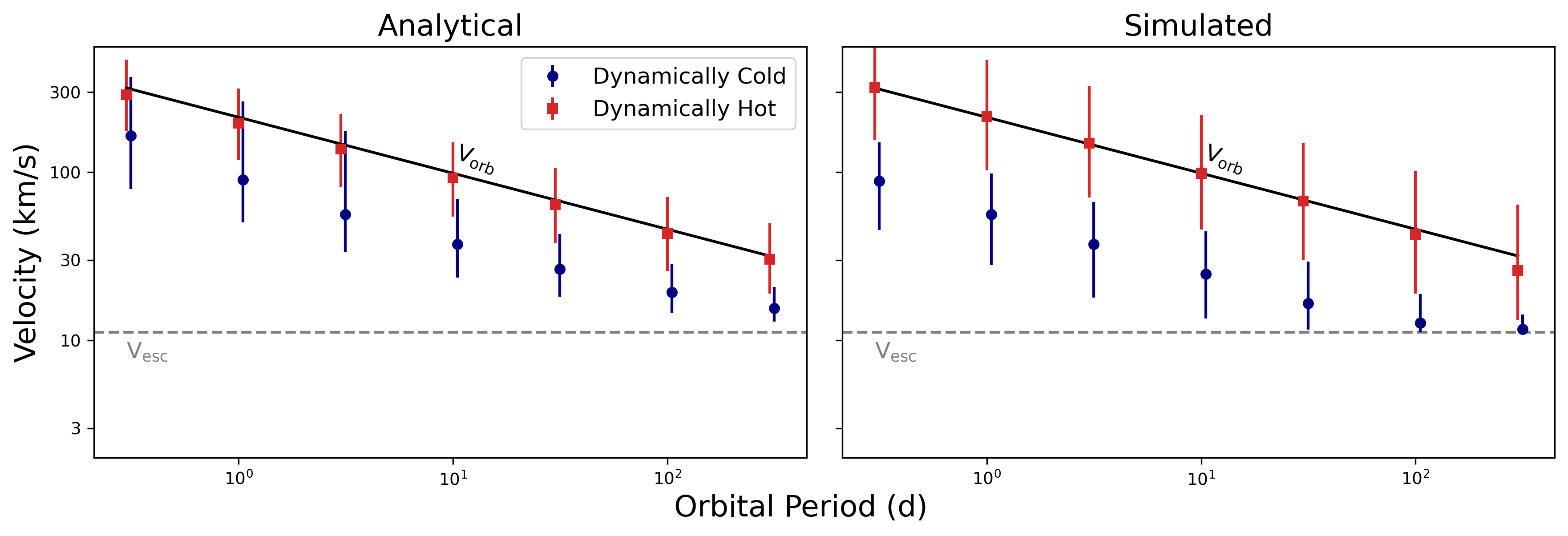}
    \caption{The period vs. $v_\mathrm{imp}$ for our analytic (left) compared to our simulated (right) calculations assuming a host star mass of 1\,M$\oplus$ and a planet mass of 1\,M$_\oplus$.  The red and blue points indicate the median (and $\pm 1\sigma$ values) of the impact velocity assuming a dynamically hot and cold disk, respectively.  A solid black line shows the orbital velocity as a function of period, and the escape velocity associated with the planet mass of 1\,M$_\oplus$ is shown as a dashed gray line.}
    \label{fig:vimp_per_circ_vs_ecc}
\end{figure*}

Next, we can compare these predictions with our simulation results.  To do so, we calculated the expected theoretical distribution of impact velocities given these equations. For each orbital period considered in our simulations, we calculated the impact velocities for $10^7$ impactor bodies with Keplerian orbital elements randomly generated the same way as they were in Section~\ref{ssec:coll} except for $\omega_i$, which is randomly set to one of the four possible values for $\omega_t$ listed in Equation~\ref{eqn:omega_condition}. The black points in Figure~\ref{fig:vimp_per_circ_vs_ecc} show the velocity distribution as a function of orbital period for these impact calculations.  Overall, our analytically derived impact velocity distributions seem to agree in terms of order-of-magnitude with those from our numerical results. Any differences (such as the roughly 50\% higher impact velocities in the dynamically-cold case) are likely due to the fact that our analytical calculations model the impactor and target as infinitesimally small objects.  The actual finite size of these objects likely allow for additional low-velocity grazing collisions on orbits that do not mathematically intersect each other.  In addition, as our simulations pre-set $\omega_t$ such that the planets are guaranteed to impact one another- therefore, the $\omega$ distribution may be different than the underlying distribution in our simulations. 


While this analytic solution is useful for getting a quick estimate of the impact velocity of planetesimals given their orbital periods, eccentricities, and inclinations, it only describes the situation in which the orbits of two planets perfectly intersect- that is, a situation where $b=0$.  Given the fact that our simulation results in Section~\ref{ssec:coll} showed that the vast majority of planet impacts occur at high values of $b$, we cannot fully abandon the non-analytic method when studying the typical conditions of planetary impacts.

\section{Discussion}
\label{sec:discussion}

\subsection{Mercury}
\label{ssec:mercury}

We can investigate the formation pathway for a Mercury by re-running our velocity simulations while inputting Mercury's orbital period and host mass.  We note that our results are sensitive to both our input initial eccentricity/inclination distributions, as well as the mass of the target body.  Therefore, without an \textit{a priori} understanding of the mass distribution of bodies in the protoplanetary disk, we cannot determine an absolute formation probability.


A wide variety of different target masses can collide to produce bodies with Mercury-like $m_{lr}$ values, as $m_{lr}$ is a function of both the $m_t$ and $v_\mathrm{imp}$ (which is, in and of itself, dependent on $m_t$ due to gravitational focusing).  A high-velocity fragmenting collison between two high-mass bodies can produce an object with a low, Mercury-like mass, but a low-velocity merging collision between two sub-Mercuries can produce a remnant with a similar mass.  However, these two different scenarios would result in extremely different outcomes- a low-velocity merging collision would result in a Mercury with a low $CMF$ value, while a high-velocity fragmentary collision would create a Mercury with a high $CMF$ value.  Therefore, when considering whether or not Mercury can be formed via collisions, it is necessary to consider which types of collisions can produce remnants with \textit{both} its mass and $CMF$.

We estimated the likelihood of a collision forming Mercury by following a similar process as outlined in Section~\ref{sssec:calc_vimp}.  We fixed the host star mass at 1\,$M_\odot$ and planet period at $88$d, and then performed $10^9$ simulations each for a grid of target masses between 1-20\,M$_\oplus$ in both the dynamically cold and dynamically hot cases. For each impact, we used the \citetalias{Dou2024} relations to determine the mass and CMF of the resulting body and then compared it to Mercury.  We used Mercury's literature mass of $0.055\,M_\oplus$ \cite{Anderson1987} and estimated a CMF of Mercury using \texttt{exopie} \citep{Plotnykov2024}, which estimates CMFs utilizing the SUPEREARTH model described in \cite{Valencia2006} and \cite{Plotnykov2020}.  Using the mass of Mercury from \cite{Anderson1987} and the radius from \cite{Perry2011}, we estimated a Mercury CMF of approximately $0.64\,\pm\,0.02$.

\begin{figure*}
    \centering
    \includegraphics[width=1\linewidth]{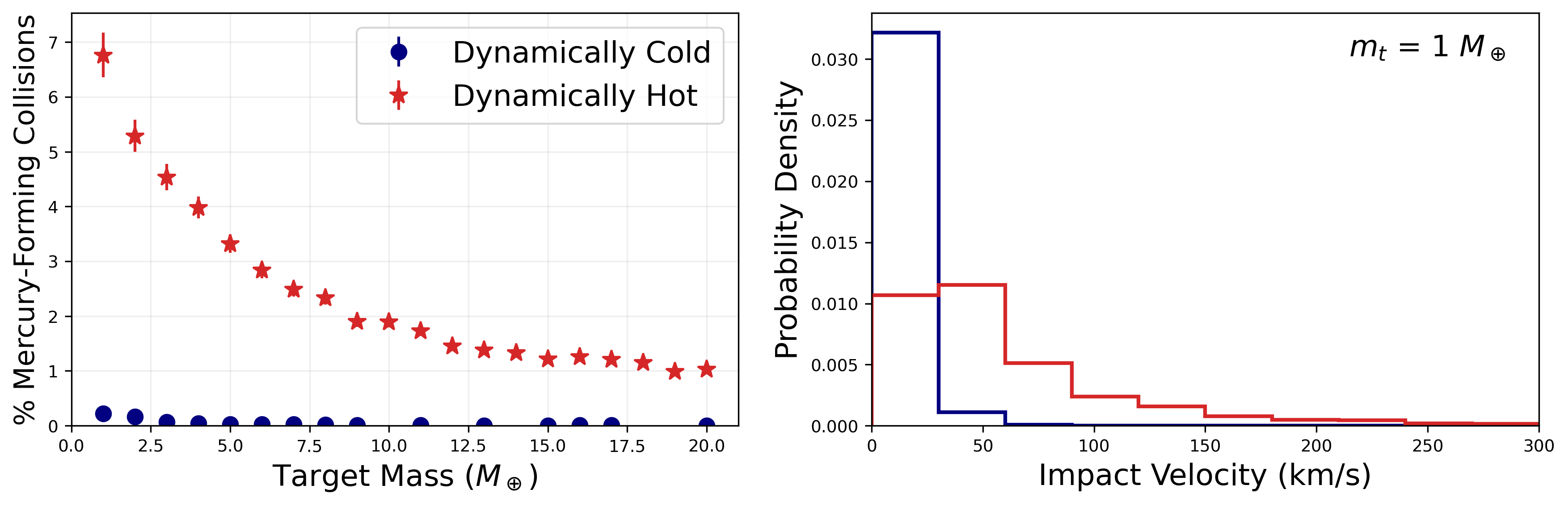}
    \caption{The velocity (left) and CMF (right) distribution of our simulated impacts to form a Mercury-like planet.}
    \label{fig:impacts_mercury}
\end{figure*}

Figure~\ref{fig:impacts_mercury} shows the velocity distribution of the simulated impacts, as well as the proportion of impacts that were able to produce Mercury-like bodies as a function of the target mass $m_t$.  We calculated the probability of an impact forming a planet as dense as (or denser than) and as large as (or larger than) Mercury using binomial statistics, following similar logic as that used in \cite{Howard2012}, estimating the $\pm$1$\sigma$ errors on the formation probability using the cumulative binomial distribution.  Overall, we found that collisions between low-mass target bodies were more likely to produce a Mercury-like planet than collisions between high-mass target bodies.  This is due to the fact that lower-mass bodies have smaller catastrophic disruption thresholds, meaning that lower-mass targets can be disrupted more easily by low-velocity collisions (which are more typical in planets with longer orbital periods).  We found that the probability of a collision creating a planet as dense as or denser than Mercury peaks around 0.2\% in the dynamically-cold case and 7\% in the dynamically-hot case.  

The formation probability appears quite high in the dynamically-hot scenario, but we caution that our estimate of a dynamically-hot disk (in which there is a uniform distribution of planetesimal eccentricities) represents an upper limit.  In addition, considering the fact that both of our disk simulations are likely optimistic (they assume equal-mass impactors, which are the most energetic possible setup), it is likely even less probable for a Mercury-like planet to form collisionally at an orbital period of 88d.

Our results in the dynamically-cold scenario largely agree with those from \cite{Franco2022}, which found it difficult to form Mercury with a single giant impact (finding that it occurred in less than 1\% of collisions), and theorized that the planet may have therefore formed through multiple mantle-stripping impacts.  However, we do note that \cite{Franco2025} was able to successfully re-create a Mercury-like body as the result of a single collision.  However, the \cite{Franco2025} simulations considered a grazing hit-and-run impact scenario in which the body that eventually becomes Mercury was lower-in-mass, making it effectively the second largest fragment.  As the formulae from \cite{Reinhardt2022} and \cite{Dou2024} track the mass of the largest fragment, it is unsurprising that we were unable to reproduce the \cite{Franco2025} results.  

From our own calculations, a single-impact formation mechanism for Mercury would be more likely if the planet formed closer to the sun and then migrated outwards after formation, but there is no theoretical evidence for such a mechanism occurring on Mercury.  Given the degree of fine-tuning necessary in order to produce Mercury with our models, we conclude that it is difficult, but not impossible, to form Mercury collisionally in the dynamically-cold scenario.  However, if the protosolar disk featured large numbers of planetesimals with extreme eccentricities, Mercury is far easier to produce.  This dichotomy demonstrates that the density and orbital distribution of exoplanets may be able to provide some insight into the dynamical history of their protostellar disks.

The dynamically-cold scenario agrees with the observation of volatiles on Mercury's surface \cite{Peplowski2011}, which has also led other groups to hypothesize that it was not formed via an impact event at all, but rather due to a primordial concentration of iron in the protoplanetary disk \citep{Johansen2022}.  However, the exact mechanism for this concentration is uncertain.  \cite{Bogdan2023} expects an iron concentration around the Curie temperature of iron (1000-1200\,K) and \cite{Mah2023} anticipates an iron concentration around the evaporation temperature of iron ($\approx$1350\,K).  In addition, the accumulation of surface volatiles may not necessarily be in conflict with a collisional history for Mercury \citep{Ebel2018}.
The study of additional high-density exoplanets may be able to provide us with additional evidence regarding the primordial vs. collisional iron concentration debate.

\subsection{Exoplanets}

One way to test this mechanism is to estimate the CMFs of the observed exoplanet population and determine if these are consistent with collisional formation.

To check this, we downloaded the exoplanet data with $>3\sigma$ mass and radius precision from the NASA exoplanet archive as of 17 November 2025 and calculated the CMFs for each planet with measured masses and radii using \texttt{exopie}. The SUPEREARTH model utilized in \texttt{exopie} is defined for target masses ranging from roughly 0.01 --- 19.95\,M$_\oplus$, so we excluded all planets with masses falling outside of that range.  We note that most planets with masses $>$20\,M$_\oplus$ tend to have lower densities consistent with thick atmospheric envelopes, so removing these targets is unlikely to exclude a substantial number of high-density super-Earths.

After eliminating planets with masses outside of the model's range, we identified planets with densities consistent with CMFs that were at least 2$\sigma$ higher than that of the Earth.  These planets, as well as their parameters, are listed in Table~\ref{tab:exo_cmf}.  We then used the literature values of the planetary periods and stellar masses to estimate the likelihood of forming a planet with its given median $CMF$ or higher for each planet using the same technique we described in Section~\ref{ssec:mercury}.  

\begin{table*}
    \centering
    \begin{tabular}{|ccccccccc|}
    \hline
         Name & Mass ($M_\oplus$) & Radius  ($R_\oplus$) & Period (d) & $M_\star$ ($M_\odot$) & Ref. & CMF & \% Collisions (Cold) & \% Collisions (Hot)  \\
         \hline
         Mercury & $0.055$        & $0.38$               & 88         & 1                        & $a$, $b$ &  $0.64\,\pm\,0.02$  &$0.22^{+0.10}_{-0.07}$\% & $6.75^{+0.41}_{-0.40}$\%  \\
         GJ\,367\,b& $0.633\pm0.050$  & $0.699 \pm 0.024 $ & $0.32$ & $0.46$ & $c$ & $0.94^{+0.04}_{-0.06}$ &$0.87\pm 0.03$\% & $2.82\pm 0.03$\% \\
         HD\,137496\,b& $4.04\pm0.55$  & $1.31^{+0.06}_{-0.05} $ & $1.62$ & $1.04$ & $d$ & $0.69^{+0.16}_{-0.19}$ & $0.35\pm 0.02$\% & $2.88\pm 0.05$\% \\
         Kepler\,107\,c& $10.0\pm2.0$  & $1.597 \pm 0.026 $ & $4.90$ & $1.24$ & $e$ & $0.76^{+0.15}_{-0.18}$ & $<0.03$\% & $<0.01$\% \\
         Kepler\,1972\,b& $2.02^{+0.56}_{-0.62}$  & $0.802^{+0.042}_{-0.041}$ & $7.55$ & $1.12$ & $f$ & $0.91^{+0.06}_{-0.21}$ & $0.014^{+0.006}_{-0.005}$\% & $0.73\pm 0.03$\% \\
         Kepler\,1972\,c& $2.11^{+0.59}_{-0.65}$  & $0.868^{+0.051}_{-0.050}$ & $11.32$ & $1.12$ & $f$ & $0.89^{+0.08}_{-0.23}$ & $0.015^{+0.007}_{-0.005}$\% & $0.62\pm 0.03$\% \\
         Kepler\,406\,b& $6.35\pm1.4$  & $1.43 \pm 0.3 $ & $2.43$ & $1.07$ & $g$ & $0.76^{+0.14}_{-0.21}$ & $<0.0002$\% & $0.000024^{+0.00002}_{-0.00001}$\% \\
         \hline
    \end{tabular}
    \caption{The planet parameters (as well as their sources) for the high-density objects considered in this study.  The CMF column describes the CMF inferred using \texttt{exopie} and the rightmost  two columns describe the percentage of simulated impacts that were able to produce a planet of its median CMF and mass or higher in the dynamically-cold (left) and hot (right) scenarios.  The references are as follows: $a$) \protect\cite{Anderson1987}, \protect\cite{Perry2011}, $c$) \protect\cite{Goffo2023}, $d$) \protect\cite{AzevedoSilva2022}, $e$) \protect\cite{Bonomo2023}, $f$) \protect\cite{Leleu2022}, $g$) \protect\cite{Marcy2014}.}
    \label{tab:exo_cmf}
\end{table*}

Overall, we found that it is difficult (with a $<1\%$ probability) to produce the majority of high-density exoplanets with collisional means.   Our calculations indicate that it is extremely difficult ($<0.03\%$) to form the high-mass planets Kepler\,107\,c and Kepler\,406\,b even in the dynamically-hot scenario, which may mean 1) these planets formed via some non-collisional mechanism, 2) their masses and/or radii are inaccurate, or 3) they formed as a result of a collision between targets that are not accurately captured by our methods (such as planets with masses greater than 20\,$M_\oplus$).  We also struggled to form the longer-period Kepler\,1972\,b and c with our simulations in the dynamically-cold case, with formation percentages $<0.02\%$, but these planets become substantially easier to form in the dynamically-hot disk ($\approx 1$\%).  However, we note that the dynamically-hot disk is likely far more extreme than a realistic protoplanetesimal population, and likely produces an unrealistically high formation percentage.  We do note that a closer examination of Kepler-1972\,b and Kepler-1972\,c \citep{Leleu2022} shows that their median literature densities are substantially higher than that of pure iron, which cannot be explained by our collision-based model, which assumes all planets are composed purely of iron cores and silicate mantles.  While there are some theoretical mechanisms that can create planets denser than iron \citep[for example, the stripped giant core model from][]{Mocquet2014}, we encourage additional observations on the Kepler-1972 system to confirm the density of its planets before invoking such exotic scenarios.These results largely agree with those from \cite{Cambioni2025}, which found that large, high-density exoplanets are far more common than can be explained with collisional mechanisms.

However, of the remaining planets, GJ\,367\,b and HD\,137496\,b hold promise for collisional formation mechanisms, with collisional formation percentages around $0.5-3\%$ in the dynamically cold and hot scenarios.  If we consider the dynamically-cold scenario to be more realistic than the hot scenario, GJ\,367\,b is the easiest planet to form collisionally, making it a promising candidate for collisional formation.  It may be no coincidence that one of the highest-CMF exoplanets observed has one of the lowest masses (0.6\,M$_\oplus$) and shortest orbital periods (0.3d).   It is substantially easier to form a a GJ\,367\,b-like planet in a dynamically-cold disk with a single collision than Mercury, despite its larger mass and smaller host star.  We do note, however, that there have been other proposed mechanisms for the high density of GJ\,367\,b, as \cite{Goffo2023} also hypothesized that GJ\,367\,b's high density could be a result of the extreme radiation it is exposed to on its short-period orbit around its host star.  However, the exact mechanism for this process has not been studied.  

Interestingly, unlike Mercury, we also found that our studied high-density exoplanets are more easily formed with high target masses as opposed to low target masses.  This may be a result of Mercury's extremely small size- two very low-mass objects can collide and produce a remnant with the mass of Mercury, but far larger objects are required to make remnants as large as Kepler\,107\,c, which requires target masses $>5\,M_\oplus$ simply to form a remnant with its mass of $M_p\,=\,10\,M_\oplus$, and even larger targets to produce an iron-rich body that still retains a high mass even after the erosion of mantle material.  Similarly, GJ\,367\,b, requires target masses $\gtrsim \, 5\,M_\oplus$ to form, and the formation chance of HD\,137496\,b increased with $m_t$ all the way to the upper limit of our mass grid ($m_t=20\,M_\oplus$).  A more careful accounting of the realistic ranges of planetesimal masses in the protoplanetary disk therefore may provide additional insight into the probability of these planets forming.  If a planet can only form via a collision between two bodies that are rare or unphysical in terms of mass, we should consider a collisional formation mechanism to be unlikely for that body.

Given the small number of observed super-Mercuries, the low percentage chance of formation for the other targets does not necessarily mean they were not formed via collisions.  Super-Mercuries, with their high masses, should theoretically be easier to characterize than lower-mass planets, which have smaller RV signals, so their observed rarity is unlikely to be a result of observational bias and could just be reflective of how difficult these planets are to form.  A statistical study of the completeness of the high-density planet sample, combined with a more careful accounting for initial disk conditions in our simulations, would be necessary to determine whether or not our observed numbers are consistent with the collisional mechanisms.  However, such work is beyond the scope of this paper.  The detection and characterization of additional sub-Earths will also improve our ability to determine whether or not a collisional formation mechanism agrees or disagrees with our data.

\section{Summary and Conclusions}
\label{sec:summary_conclusions}

There are many hypotheses that explain the formation of high-density exoplanets, including the possibility that they form as the result of collisional stripping of a differentiated planet's mantle.  However, to evaluate the likelihood of this hypothesis, it will be necessary to determine if there are any observable signatures of collisional planet formation, especially when studying systems exterior to the solar system.

In this paper, we showed that collisionally-produced planets can have higher CMF values at low masses and short orbital periods.  Lower-mass super-Mercuries are easier to produce than high-mass super-Mercuries because it is easier to strip the mantles of low-mass planets, and short-period super-Mercuries are more likely to form because these bodies are more likely undergo high-energy impacts.  While there are complicating factors (such as the initial mass, eccentricity, and inclination distributions of the planetesimals), these are unlikely to affect the fundamental scaling of these relationships.

We also derived the analytic formula that describes the scaling between the impact velocity and the Keplerian orbital elements of the impactors, and found that the impact velocity is directly proportional to the orbital velocity.  In addition, when the impactor and target have similar semi-major axes, the impact velocity is directly proportional to the eccentricity (in low-inclination scenarios) and the inclination (in low-eccentricity scenarios) of the impactor, as has been discussed in works like \cite{Lissauer1993}. 

We then studied the likelihood of formation of various observed high-density planets.  We found it difficult to produce Mercury at its modern-day orbit (in general agreement with other publications), but found that it is relatively easy to form a GJ\,367\,b-like planet on a short orbital period.  Similarly to \cite{Cambioni2025}, we found that it is typically very difficult to form high-density super-Earths, indicating either that our observations of these planets are biased or that high-density super-Earths may form via some non-collisional means.  However, with the current small sample size of confirmed super-dense exoplanets (with only six exoplanets with $>2\sigma$ Earth CMFs), it is still possible that this mechanism could explain their formation.  It will be necessary to discover and characterize additional super-Mercuries, as well as evaluate the observational biases of this population, before it is possible to search for evidence of the correlations associated with collisional planet formation.

\section*{Acknowledgements}

We thank the anonymous reviewer for their constructive comments on this paper.

DZS acknowledges funding support from JWST GO 5959, which was provided by NASA through a grant from the Space Telescope Science Institute.
This research has made use of NASA's Astrophysics Data System Bibliographic Services.

\textit{Software:} Exopie \citep{Plotnykov2024}, Matplotlib \citep{matplotlib}, Numpy \citep{numpy}, Pandas \citep{mckinney-proc-scipy-2010, reback2020pandas}, Scipy \citep{2020SciPy-NMeth}, Radvel \citep{Fulton2017}

\textit{Facilities:} NASA Exoplanet Archive

\section*{Data Availability}

The data used in this article are included in the NASA Exoplanet Archive, which is accessible \href{https://exoplanetarchive.ipac.caltech.edu/index.html}{online}. 

The software utilised in this paper are publicly available and are described and cited in the Acknowledgements.

The additional data products generated as a result of the simulations will be shared on reasonable request to the corresponding author.



\bibliographystyle{mnras}
\bibliography{manuscript} 




\appendix
\section{The Position and Velocity of an Arbitrary Body}
\label{app:pv}

In \cite{Schwarz2017}, the position $p$ of a target in x, y, and z coordinates is as follows:

\begin{equation}
\begin{aligned}
p_x = a \bigg(1-e\,\mathrm{cos}\big(E(t)\big)\bigg) \times \\\Bigg(\mathrm{cos\big(f(t)\big)} \bigg(\mathrm{cos}\big(\omega\big) \mathrm{cos}\big(\Omega\big) -\\\mathrm{sin}\big(\omega\big)\mathrm{cos}\big(i\big)\mathrm{sin}\big(\Omega\big)\bigg) -\\ \mathrm{sin\big(f(t)\big)} \bigg( \mathrm{sin}\big(\omega\big)\mathrm{cos}\big(\Omega\big)+\\\mathrm{cos}\big(\omega\big)\mathrm{cos}\big(i\big)\mathrm{sin}\big(\Omega\big)\bigg)\Bigg),
\end{aligned}
\end{equation}

\begin{equation}
\begin{aligned}
p_y = a \bigg(1-e\,\mathrm{cos}\big(E(t)\big)\bigg) \times \\\Bigg(\mathrm{cos\big(f(t)\big)} \bigg(\mathrm{cos}\big(\omega\big) \mathrm{sin}\big(\Omega\big) +\\\mathrm{sin}\big(\omega\big)\mathrm{cos}\big(i\big)\mathrm{cos}\big(\Omega\big)\bigg) + \\\mathrm{sin\big(f(t)\big)} \bigg( \mathrm{cos}\big(\omega\big)\mathrm{cos}\big(i\big)\mathrm{cos}\big(\Omega\big)-\\\mathrm{sin}\big(\omega\big)\mathrm{sin}\big(\Omega\big)\bigg)\Bigg),
\end{aligned}
\end{equation}

\noindent and:

\begin{equation}
\begin{aligned}
\label{eqn:pz}
    p_z = a \bigg(1-e\,\mathrm{cos}\big(E(t)\big)\bigg) \times \\\Bigg(\mathrm{cos\big(f(t)\big)} \mathrm{sin}\big(\omega\big) \mathrm{sin}\big(i\big)  +\\ \mathrm{sin\big(f(t)\big)} \mathrm{cos}\big(\omega\big)\mathrm{sin}\big(i)\Bigg).
\end{aligned}
\end{equation}

Meanwhile, the velocity $v$ is as follows:
\begin{equation}
\begin{aligned}
    v_x = \frac{\sqrt{GM_\star a}}{a \bigg(1-e\,\mathrm{cos}\big(E(t)\big)\bigg)} \times \\ \Bigg(-\mathrm{sin\big(E(t)\big)} \bigg(\mathrm{cos}\big(\omega\big) \mathrm{cos}\big(\Omega\big) - \\\mathrm{sin}\big(\omega\big)\mathrm{cos}\big(i\big)\mathrm{sin}\big(\Omega\big)\bigg) - \\ \sqrt{1-e^2}\mathrm{cos\big(E(t)\big)} \bigg( \mathrm{sin}\big(\omega\big)\mathrm{cos}\big(\Omega\big)+ \\\mathrm{cos}\big(\omega\big)\mathrm{cos}\big(i\big)\mathrm{sin}\big(\Omega\big)\bigg)\Bigg),
\end{aligned}
\end{equation}

\begin{equation}
\begin{aligned}
v_y = \frac{\sqrt{GM_\star a}}{a \bigg(1-e\,\mathrm{cos}\big(E(t)\big)\bigg)}  \times\\  \Bigg(-\mathrm{sin\big(E(t)\big)} \bigg(\mathrm{cos}\big(\omega\big) \mathrm{sin}\big(\Omega\big) +\\\mathrm{sin}\big(\omega\big)\mathrm{cos}\big(i\big)\mathrm{cos}\big(\Omega\big)\bigg) +\\ \sqrt{1-e^2}\mathrm{cos\big(E(t)\big)} \bigg( \mathrm{cos}\big(\omega\big)\mathrm{cos}\big(i\big)\mathrm{cos}\big(\Omega\big)-\\\mathrm{sin}\big(\omega\big)\mathrm{sin}\big(\Omega\big)\bigg)\Bigg),
\end{aligned}
\end{equation}

\noindent and:

\begin{equation}
\begin{aligned}
v_z = \frac{\sqrt{GM_\star a}}{a \bigg(1-e\,\mathrm{cos}\big(E(t)\big)\bigg)}  \times\\\Bigg(-\mathrm{sin\big(E(t)\big)} \mathrm{sin}\big(\omega\big) \mathrm{sin}\big(i\big)  +\\ \sqrt{1-e^2}\mathrm{cos\big(E(t)\big)} \mathrm{cos}\big(\omega\big)\mathrm{sin}\big(i)\Bigg).
\end{aligned}
\end{equation}

As the eccentric anomaly $E$ by definition is a function of the mean anomaly $f$ and eccentricity $e$, we can describe the position of and velocity of a particle at a time $t$ as a function of $a$, $e$, $i$, $f(t)$, $\omega$, and $\Omega$.

\section{Approximating $v_\mathrm{imp}$}
\label{app:vimp}

When the impactor has a similar semi-major axis as the target (that is, $a_i\approx a_t$), $v_\mathrm{imp}$ dramatically simplifies to:

\begin{equation}
\label{eqn:vimp_simplified}
    v_\mathrm{imp} = v_{orb,t} \sqrt{ \Bigg(2 \mp 2  \mathrm{cos}\big(i_i\big)  \mathrm{sin} \big(\omega_i\big) \Bigg)},
\end{equation}

\noindent where the ``minus'' solution corresponds to an even value of $n$ and the ``plus'' solution corresponds to an odd value of $n$.

This can be further simplified by considering the definition of $\omega_i$:

\begin{equation}
    \omega_i =-2\,\mathrm{tan}^{-1}\Bigg(\frac{\sqrt{1+e_i}}{\sqrt{1-e_i}\,} \mathrm{tan}\bigg(\frac{1}{2} \mathrm{cos}^{-1}\bigg(\frac{a_i-a_t + a_t e_t\,\mathrm{cos} \big(E_t\big)}{e_i a_i}\bigg) \bigg)\Bigg) + n\pi.
\end{equation}

If $e_t=0$ and $a_t\approx a_i$:

\begin{equation}
\begin{aligned}
    \omega_i = -2\,\mathrm{tan}^{-1}\Bigg(\frac{\sqrt{1+e_i}}{\sqrt{1-e_i}\,} \mathrm{tan}\bigg(\frac{1}{2} \mathrm{cos}^{-1}\big(0\big) \bigg)\Bigg) + n\pi \\= - 2\,\mathrm{tan}^{-1}\Bigg(\frac{\sqrt{1+e_i}}{\sqrt{1-e_i}\,}\Bigg) + n\pi.
\end{aligned}
\end{equation}

For low values of eccentricity $e_i$, we can calculate an approximation for $\omega_i$ by Taylor expanding around $x=0$:

\begin{equation}
\label{eqn:texpansion_omega}
    \omega_i \approx n\pi - \frac{\pi}{2} - e_i - \frac{e_i^3}{6} + O(e^5)
\end{equation}

We will omit the factors on the order of $e_i^5$, under the assumption that such a value is negligibly small for small eccentricity values.

Now, we see that $\omega_i$ is around $-\pi/2$ for low values of eccentricity.  We Taylor expanded around $\omega = n\pi -\pi/2$ to estimate sin($\omega_i$):

\begin{equation}
\begin{aligned}
    \mathrm{sin}(\omega_i) =  -\cos(n\pi)  + \frac{1}{2} \cos(n \pi)\bigg(\omega_i + \frac{\pi}{2} - n\pi\bigg)^2 \\-\frac{1}{24}\cos(n \pi)\bigg(\omega_i + \frac{\pi}{2} - n\pi\bigg)^4 + O\Bigg(\bigg(\omega_i+\frac{\pi}{2} - n\pi\bigg)^6 \Bigg),
\end{aligned}
\end{equation}

\noindent and, with Equation~\ref{eqn:texpansion_omega}:

\begin{equation}
\begin{aligned}
    \mathrm{sin}(\omega_i) = -\cos(n\pi) + \frac{1}{2}\cos(n\pi) \bigg(- e_i - \frac{e_i^3}{6}\bigg)^2 \\- \frac{1}{24}\cos(n\pi)\bigg(- e_i - \frac{e_i^3}{6}\bigg)^4 + O\Bigg(\bigg(- e_i - \frac{e_i^3}{6}\bigg)^4 \Bigg)
\end{aligned}
\end{equation}

If we remove all factors of order $e^4$ and higher on the grounds that they are negligible, we find:

$$
\mathrm{sin}(\omega_i) \approx \begin{cases}
    -1 + \frac{1}{2} e_i^2 & \text{if } ~n~\mathrm{even}\\
    1 - \frac{1}{2} e_i^2 & \text{if } ~n~\mathrm{odd}.\\
\end{cases}
$$

Similarly, if we assume the inclination is low, we can expand cos($i_i$):

\begin{equation}
    \mathrm{cos}(i_i) \approx 1 - \frac{i_i^2}{2} + O\big(i_i^4\big).
\end{equation}

Substituting these values back into Equation~\ref{eqn:vimp_simplified} and assuming that factors of order $i_i^4$ are negligibly small:

\begin{equation}
    v_\mathrm{imp} \approx v_{orb,t} \sqrt{ \Bigg(2 + 2  \bigg(-1 + \frac{1}{2} e_i^2\bigg) \bigg(1 - \frac{i_i^2}{2} \bigg) \Bigg)}
\end{equation}

\begin{equation}
    v_\mathrm{imp} \approx v_{orb,t} \sqrt{ \Bigg(2 + 2  \bigg(-1 +\frac{1}{2}e_i^2 +\frac{1}{2}i_i^2 -\frac{1}{4}e_i^2i_i^2  \bigg)}
\end{equation}

We found (omitting the $i^2e^2$ factor, which is likely also negligibly small):

\begin{equation}
    v_\mathrm{imp} \approx v_{orb,t} \sqrt{e_i^2 + i_i^2 }
\end{equation}

We note that this expression, which simplifies to $v_\mathrm{imp} = v_{orb,t}e_i$ when $i_i$ is small and $v_\mathrm{imp} = v_{orb,t}i_i$ when $e$ is small, is identical to that in \cite{Lissauer1993}.


\bsp	
\label{lastpage}
\end{document}